\documentclass[a4paper,fleqn,usenatbib]{mnras}

\usepackage{newtxtext,newtxmath}

\usepackage[T1]{fontenc}
\usepackage{ae,aecompl}

\usepackage{graphicx}	
\usepackage{amsmath}	
\usepackage{amssymb}	

\newcommand{\aref}[1]{\hyperref[#1]{Appendix~\ref{#1}}}

\usepackage{color}
\usepackage[normalem]{ulem}

\definecolor{darkgreen}{rgb}{0.13, 0.55, 0.13}

\title[Fundamental test for feedback recipes]{\vspace{-4mm}A fundamental test for stellar feedback recipes in galaxy simulations\vspace{-3mm}}

\author[Y.~Fujimoto et al.]
{Yusuke Fujimoto,$^{1}$\thanks{E-mail: yusuke.fujimoto@anu.edu.au}
M{\'e}lanie Chevance,$^{2}$
Daniel T.~Haydon,$^{2}$
Mark R.~Krumholz,$^{1,3}$
\newauthor
and J.~M.~Diederik Kruijssen$^{2}$
\\
$^{1}$Research School of Astronomy \& Astrophysics, Australian National University, Canberra, Australian Capital Territory 2611, Australia\\
$^{2}$Astronomisches Rechen-Institut, Zentrum f{\"u}r Astronomie der Universit{\"a}t Heidelberg, M{\"o}nchhofstra{\ss}e 12-14, 69120 Heidelberg, Germany\\
$^{3}$Centre of Excellence for Astronomy in Three Dimensions (ASTRO-3D), Australia
\vspace{-3mm}}

\date{Accepted XXX. Received YYY; in original form ZZZ\vspace{-2mm}}

\pubyear{2019}

\begin{document}
\label{firstpage}
\pagerange{\pageref{firstpage}--\pageref{lastpage}}
\maketitle

\begin{abstract}
Direct comparisons between galaxy simulations and observations 
that both reach scales $\lesssim 100$ pc
are strong tools to investigate the cloud-scale physics of star formation and feedback in nearby galaxies. Here we carry out such a comparison for hydrodynamical simulations of a Milky Way-like galaxy, including stochastic star formation, $\mathrm H\textsc{ii}$ region and supernova feedback, and chemical post-processing at 8 pc resolution. Our simulation shows excellent agreement with almost all kpc-scale and larger observables, including total star formation rates, radial profiles of CO, H\textsc{i}, and star formation through the galactic disc, mass ratios of the ISM components, both whole-galaxy and resolved Kennicutt-Schmidt relations, and giant molecular cloud properties. However, we find that our simulation does not reproduce the observed de-correlation between tracers of gas and star formation on $\lesssim 100$ pc scales, known as the star formation `uncertainty principle', which indicates that observed clouds undergo rapid evolutionary lifecycles. We conclude that the discrepancy is driven by insufficiently-strong pre-supernova feedback in our simulation, which does not disperse the surrounding gas completely, leaving star formation tracer emission too strongly associated with molecular gas tracer emission, inconsistent with observations. This result implies that the cloud-scale de-correlation of gas and star formation is a fundamental test for feedback prescriptions in galaxy simulations, one that can fail even in simulations that reproduce all other macroscopic properties of star-forming galaxies.
\end{abstract}

\begin{keywords}
hydrodynamics -- 
methods: numerical -- 
ISM: clouds -- 
ISM: kinematics and dynamics -- 
galaxies: evolution -- 
galaxies: ISM
\vspace{-3mm}
\end{keywords}





\section{Introduction}

Comparison with observation is essential to developing theoretical and numerical models that can accurately describe real galaxies. Numerical simulations bring significant benefits by providing insight into the time evolution of gas dynamics and star formation within a galaxy, something that is difficult to constrain via observations because they provide only static snapshots. However, for us to have confidence in these insights we must rigorously check the simulations against as many observational constraints as possible. Thus close cooperation between theory and observation is crucial for a complete understanding of star formation and feedback in a galaxy.

One important observational constraint is the star formation rate (SFR). Galactic-scale star formation appears to obey a power-law relation between the gas surface density and the surface density of the star formation rate, the so-called Kennicutt-Schmidt relation \citep[][]{Schmidt1959, Kennicutt1989}. This correlation holds not just for averages of local galaxies, but also for $\sim\mbox{kpc}$-sized patches in nearby spiral galaxies \citep[e.g.][]{BigielEtAl2008, SchrubaEtAl2011, KennicuttEvans2012, LeroyEtAl2013}, and for whole galaxies out to high redshift \citep[e.g.][]{DaddiEtAl2010, GenzelEtAl2010, TacconiEtAl2013}. The galactic star formation relation has been used extensively to calibrate stellar feedback recipes in galaxy simulations. Supernova (SN) explosions are the most important source of feedback in simulations of galaxy formation and evolution, including simulations on cosmological scales \citep[e.g.][]{CenOstriker1992, SpringelHernquist2003, ScannapiecoEtAl2006} all the way down to parsec scales \cite[e.g.][]{JoungMacLow2006, KimKimOstriker2011, KimOstriker2015}. For example, it has been known for many years that simply depositing $E_{\rm SN} = 10^{51}$ erg per SN as thermal energy in the location of a SN explosion does not produce feedback strong enough for star formation rates to match observed ones, because the energy is quickly radiated away due to low resolution (typically 10-1000 pc), the so-called overcooling problem \citep{Katz1992}. 
This problem can be overcome by reaching resolutions high enough to capture the Sedov-Taylor phase of supernova remnant expansion \citep[e.g.,][]{ForbesEtAl2016,HuEtAl2016,Hu2019,EmerickEtAl2019}, but at present this is technically possible only in simulations of dwarf galaxies, where the low overall galaxy size and the low density (leading to extended Sedov-Taylor phases) eases the resolution requirement.
To fix this problem 
in simulations of larger galaxies
, authors have turned to various types of momentum or mixed energy/momentum SN feedback recipes \citep[e.g.,][]{Kimm15a,GoldbaumKrumholzForbes2016,Hopkins18a}, and have included other forms of feedback, such as radiation pressure from massive stars and photoionization heating of $\mathrm H\textsc{ii}$ regions \citep[e.g.][]{HopkinsQuataertMurray2011, AgertzKravtsovLeitner2013}. Because these feedback processes occur on scales that are not resolved in galaxy-scale or cosmological simulations, the recipes for them are calibrated using a mix of smaller-scale simulations and direct comparisons between simulation and the observed Kennicutt-Schmidt relation.

A second important observational constraint that can be used to validate simulations is the turbulent velocity structure of the interstellar medium (ISM) and, closely related the gravitational stability or instability of the galactic disc. $\mathrm H\textsc{i}$ and CO lines in nearby disc galaxies show super-thermal velocity dispersions of $\sim 10\ \mathrm{km\ s^{-1}}$ \citep[e.g.][]{TamburroEtAl2009, IanjamasimananaEtAl2012, Caldu-Primo13a}, and the velocity dispersions are such that these galaxies' Toomre $Q$ parameters \citep{Toomre1964} are about unity, indicating that the galactic discs are marginally gravitationally stable \citep[][]{LeroyEtAl2008}. H$\alpha$ observations of higher redshift galaxies show higher velocity dispersions, but comparable Toomre $Q$ values in the star-forming parts of galaxy discs \citep[e.g.,][]{Genzel14a}. Galaxy simulations have shown that the combination of self-gravity, radiative cooling, and galactic shear motion drive turbulence into the ISM and increase the total $Q$, while stellar feedback prevents runaway fragmentation of star-forming gaseous clumps and moderates the consumption of gas, keeping $Q$ close to unity \citep{AgertzRomeoGrisdale2015, GoldbaumKrumholzForbes2015, GoldbaumKrumholzForbes2016, GrisdaleEtAl2017}. The ability of simulations to maintain discs in a state or marginal stability, and reproduce the observed velocity dispersions and their evolution with redshift \citep[e.g.,][]{Hung18a} represents another important point of observational comparison.

The properties of giant molecular clouds (GMCs) provide a third point of contact between simulations and observations. GMCs are the coldest gas in the galactic ISM and are stellar nurseries. Observations of CO line emission have revealed their spatial distribution and demographics, such as their spectrum of mass, size, and velocity dispersion, in both the Milky Way and nearby galaxies \citep[e.g.][]{SolomonEtAl1987, RosolowskyEtAl2003, HeyerEtAl2009, KodaEtAl2009, RomanDuvalEtAl2010}. Galaxy simulations have also investigated GMC formation and evolution and succeeded in reproducing many observed GMC properties \citep[e.g.][]{DobbsBonnellPringle2006, DobbsBurkertPringle2011, TaskerTan2009, TaskerWadsleyPudritz2015, FujimotoEtAl2014, FujimotoEtAl2016, GrisdaleEtAl2018}. However, they can do so only for some feedback and star formation recipes -- for example, insufficiently strong feedback leads to a GMC mass spectrum that is too top-heavy \citep[e.g.,][]{DobbsBurkertPringle2011}, while overly-efficient star formation leaves too little of the interstellar medium in the molecular phase \citep[e.g.,][]{Semenov18a} -- providing another constraint that can be used to calibrate these recipes.

While the ability of simulations to reproduce each of these observational constraints is heartening, all of them are relatively far-removed from the small-scale, time-dependent, non-equilibrium behaviour for which we need simulations the most, because observations offer little guidance. Ideally, we would like to use simulations to improve our understanding of the total life cycle of cloud-scale star formation: e.g.\ from cloud formation by galactic-scale gas dynamics and ISM physics, through clouds' dispersal by stellar feedback, to re-aggregation of clouds by compressive turbulence of the ISM from SN explosions, galactic shocks due to spiral arms and so on. There have been some numerical attempts to quantify this cycle, for example  \citet{Semenov17a, Semenov18a}, but because quantifying the timescales of GMC evolution remains an outstanding observational problem, we have for the most part lacked observational constraints that can provide direct checks on the cloud-scale physics of star formation and feedback in galaxy simulations. However, \citet{kruijssen14} and \citet{KruijssenEtAl2018} recently proposed a new statistical method, the so-called `uncertainty principle for star formation', for exactly this purpose. The basic idea behind this method is to use the scale-dependent correlation or anti-correlation between tracers of molecular gas and star formation \citep[e.g.,][]{OnoderaEtAl2010,SchrubaEtAl2010} as a function of spatial resolution to derive the evolutionary timeline of GMCs and star-forming regions. The ability of this method to directly constrain the cloud-scale processes of star formation and feedback suggests that the anti-correlation of gas and star formation at small scales, and the transition from anti-correlation to correlation as the scale increases, may provide a powerful new observational constraint on cloud-scale physics in galaxy simulations. 

In this paper, we explore this possibility using a state-of-the-art hydrodynamical galaxy simulation that includes essentially all of the features normally found in modern zoom-in cosmology or isolated galaxy simulation methods: stochastic star formation treated at a resolution sufficient to capture individual SNe, pre-SN photoionization feedback, and a uniformly-high resolution of 8 pc.\footnote{We note that 8 pc might not at first appear to be high resolution for readers who are most familiar with Lagrangian cosmological or galaxy simulations, where the convention for spatial resolutions is to quote gravitational or hydrodynamic force resolution rather than mean spatial resolution, the figure generally quoted for Eulerian simulations. To translate our resolution into Lagrangian form, note that a spatial resolution of 8 pc corresponds to a mass resolution of $18$ $\mathrm{M_\odot}$ at the mean 1 cm$^{-3}$ density of the Milky Way's ISM, or $\approx 2000$ $\mathrm{M_\odot}$ at the mean density of GMCs.} We compare this simulation to a wide range of observations, including the star formation uncertainty principle. 
While a number of previous numerical studies have reported increasing scatter in the correlation between star formation and gas tracers at smaller spatial scales \citep[e.g.,][]{KimEtal2013, HuEtAl2016}, qualitatively consistent with the observations that form the basis of the uncertainty principle, the study we present here represents the first quantitative comparison between the scale-dependent correlation found in simulations and that observed for real galaxies.
In \autoref{Methods}, we present our numerical model of a Milky Way-like galaxy. In \autoref{Results}, we describe results of analyses of our simulation and comparisons with observations. We will show that our simulation reproduces all macroscopic (i.e., $\gtrsim 1$ kpc-scale) and mesoscopic (i.e., $\gtrsim 100$ pc-scale) observational constraints but fails to reproduce the star formation uncertainty principle. We summarise our findings and discuss their implications in \autoref{Conclusions}.

\section{Methods}
\label{Methods}

We carry out our project of comparing ``observations'' of a simulation to real observations using the Milky Way-like galaxy simulation described in \citet{FujimotoKrumholzTachibana2018}. We refer readers to that paper for full details of the numerical method, and here simply summarise the most important aspects of the simulation in \autoref{Galaxy model}, followed by a discussion of our implementation of star formation and feedback in \autoref{Star formation and feedback} and then an explanation of how we carry out our chemical post-processing in \autoref{Chemical and observational post-processing}.

\subsection{Galaxy model}
\label{Galaxy model}

Our simulations follow the evolution of a Milky-Way-type galaxy using the adaptive mesh refinement code \textsc{enzo} \citep{BryanEtAl2014}. The root grid is $128^3$ in a 3D box of $(128\ \rm kpc)^3$. An additional seven levels of refinement is included, producing a minimum cell size of 7.8125 pc. We use a piecewise parabolic mesh hydrodynamics solver to follow the motion of the gas.

The gas cools radiatively to 10 K using a 1D cooling curve created from the \textsc{cloudy} package's cooling table for metals and \textsc{enzo}'s non-equilibrium cooling rates for atomic species of hydrogen and helium \citep{AbelEtAl1997, FerlandEtAl1998}. In addition to radiative cooling, the gas can also be heated via diffuse photoelectric heating in which electrons are ejected from dust grains via far-ultraviolet (FUV) photons. This is implemented as a constant heating rate of $8.5 \times 10^{-26}\ \mathrm{erg\ s^{-1}}$ per hydrogen atom uniformly throughout the simulation box. Self-gravity of the gas is also implemented.

We use initial conditions identical to those of \citet{TaskerTan2009}. The simulated galaxy is set up as an isolated disc of gas orbiting in a static background potential which represents both dark matter and a stellar disc component. The background potential is logarithmic \citep{BinneyTremaine2008} with a constant circular velocity of 200 $\mathrm{km\ s^{-1}}$ at large radii ($r > 2$ kpc). The initial gas distribution is chosen to give a constant value of the Toomre $Q$ for gravitational instability \citep{Toomre1964}. The initial gas disc is divided radially into three parts. Between radii of $r = $ 2 - 13 kpc, the gas is set so that $Q = 1$. The other regions of the galaxy, from 0 to 2 kpc and from 13 to 14 kpc, are initialized with $Q = 20$. Beyond 14 kpc, the disc is surrounded by a static, very low density medium. The initial gas mas is $8.6 \times 10^9\ \mathrm{M_{\odot}}$.

We run the simulation for 730 Myr, gradually increasing the resolution to our maximum and letting the ISM reach a statistically steady state. We then run at maximum resolution from $t=730 - 850$ Myr. Whenever we discuss time-averaged behaviour in this paper, the average is drawn from this time interval. Whenever we analyse an individual snapshot, we select the snapshot at $t=790$ Myr.

\subsection{Star formation and feedback}
\label{Star formation and feedback}

Star formation is parametrized by two choices: a threshold density at which star formation begins, and an efficiency of star formation per free-fall time in cells above that threshold. We use a resolution-dependent number density threshold for star formation of $57.5\ \mathrm{cm}^{-3}$ for $\Delta x = 8$ pc; this density is chosen so that it corresponds to the density that is Jeans unstable at our maximum resolution for the equilibrium temperature dictated by our heating and cooling processes. The second parameter is the star formation efficiency per free-fall time for gas above the density threshold. We express the SFR density in cells that exceed the threshold as
\begin{equation}
    \dot{\rho_*} = \epsilon_{\mathrm{ff}} \frac{\rho}{t_{\mathrm{ff}}}.
\end{equation}
Here, $\rho$ is the gas density of the cell, $t_{\mathrm{ff}} = \sqrt{3 \pi / 32 \mathrm{G} \rho}$ is the local free-fall time, and $\epsilon_{\mathrm{ff}} = 0.01$ is the rate parameter; this value is chosen based on extensive observational evidence (see e.g.~\citealt{KrumholzTan2007,LeroyEtAl2017,UtomoEtAl2018} and the review by \citealt{KrumholzEtAl2018}), and is also the value found in other high-resolution simulations to give approximately the correct balance of ISM phases \citep[e.g.,][]{GoldbaumKrumholzForbes2016, Semenov18a}. To avoid creating an extremely large number of star particles with the associated heavy computational costs, we impose a minimum star particle mass, $m_{\mathrm{sf}}$, and form star particles stochastically rather than spawn particles in every cell at each time-step. In this scheme, a cell forms a star particle of mass $m_{\mathrm{sf}} = 300\ \mathrm{M_{\odot}}$ with probability
\begin{equation}
    P = \left(\epsilon_{\mathrm{ff}} \frac{\rho}{t_{\mathrm{ff}}} \Delta x^3 \Delta t\right) / m_{\mathrm{sf}},
\end{equation}
where $\Delta x$ is the cell width, and $\Delta t$ is the simulation time-step. In practice, all star particles in our simulation are created via this stochastic method with masses equal to $m_{\mathrm{sf}}$. Star particles are allowed to form in the main region of the disc between $2 < r < 14$ kpc.

Within each of these particles we expect there to be a few stars massive enough to produce SN explosions and substantial ionising luminosities. We model this using the \textsc{slug} stellar population synthesis code \citep{daSilvaFumagalliKrumholz2012, KrumholzEtAl2015}. This stellar population synthesis method is used dynamically in our simulation; each star particle spawns an individual \textsc{slug} simulation that stochastically draws individual stars from the IMF, tracks their mass- and age- dependent ionising luminosities, and determines when individual stars explode as SNe.

We include stellar feedback from photoionization and SNe, following \citet{GoldbaumKrumholzForbes2016}, though our numerical implementation is very similar to that used by a number of previous authors \citep[e.g.][]{HopkinsQuataertMurray2012, AgertzKravtsovLeitner2013}. For the former, we use the total ionising luminosity $S$ from each star particle calculated by \textsc{slug} to estimate the Str\"{o}mgren volume $V_s = S/\alpha_{\rm B} n^2$, and compare with the cell volume, $V_c$. Here $\alpha_{\rm B} = 2.6\times 10^{-13}$ cm$^3$ s$^{-1}$ is the case B recombination rate coefficient, $n = \rho/\mu m_{\rm p}$ is the number density, and $\mu = 1.27$ and $m_{\rm p} = 1.67 \times 10^{-24}$ g are the mean atomic weight and the mass of a proton, respectively. If $V_s < V_c$, the cell is heated to $10^4 (V_s/V_c)$ K. If $V_s > V_c$, the cell is heated to a temperature of $10^4$ K, and then we calculate the luminosity $S_{\rm esc} = S - \alpha_{\rm B} n^2 V_c$ that escapes the cell. We distribute this luminosity evenly over the neighbouring 26 cells, and repeat the procedure.

For SN feedback, we identify particles that will produce SNe in any given time step. For each SN that occurs, we add a total momentum of $5 \times 10^5\ \mathrm{M_{\odot}}\ \rm km\ s^{-1}$ \citep{GentryEtAl2017}, directed radially outward in the 26 neighbouring cells. The total net increase in kinetic energy in the cells surrounding the SN host cell are then deducted from the available budget of $10^{51}$ erg and the balance of the energy is then deposited in the SN host cell as thermal energy. 

We include gas mass injection from stellar winds and SNe to each star particle's host cell during each time step. The mass loss rate of each star particles is calculated from the \textsc{slug} stellar population synthesis model.

\subsection{Chemical and observational post-processing}
\label{Chemical and observational post-processing}

While we do not include chemical transition from atomic gas to molecular gas in this simulation, we calculate $\mathrm{H\textsc{i}}$, $\mathrm{H_2}$, and CO mass fractions, and the associated luminosities of the H\textsc{i}-21 cm and CO $J=1\rightarrow 0$ lines, using the post-processing astrochemistry and radiative transfer code \textsc{despotic} \citep{Krumholz2014}. We use \textsc{despotic} to generate a table of cloud models where for each model we compute the H\textsc{i}, H$_2$, and CO mass fractions, and associated line luminosities, as a function of the total number density of H nuclei, $n_{\mathrm{H}}$, the column density of H nuclei, $N_{\mathrm{H}}$, and the virial parameter $\alpha_{\mathrm{vir}}$. For each model cloud, we use \textsc{despotic} to find the chemical, thermal, and statistical equilibrium state of the cloud. The chemical equilibrium calculation uses the C-O chemical network of \citet{GongEtAl2017}, whereas the thermal equilibrium calculation includes the heating by cosmic rays and the (shielding-modified) grain photoelectric effect, cooling by the lines of C~\textsc{ii}, C~\textsc{i}, O~\textsc{i}, and CO, and thermal exchange between dust and gas, and the statistical equilibrium calculation uses the escape probability formalism assuming spherical geometry. For details on how all of these processes are treated, see \citet{Krumholz2014}. We take the interstellar radiation field (ISRF) strength to be unity in \citet{Draine78a} units (1.6 in \citealt{Habing68a} units), and the primary cosmic ray ionisation rate to be $2\times 10^{-16}$~s$^{-1}$ \citep[e.g.,][]{Indriolo12a}. Since the chemical, thermal, and statistical networks are coupled (e.g., thermal equilibrium depends on CO cooling rate and thus on the statistical distribution of CO levels, chemical reaction rates depend on the temperature, etc.), the entire system is iterated to convergence simultaneously; see \citet{Krumholz2014} for details. The output of this calculation is the mass fraction in each chemical state, and the emergent luminosities of all major lines, properly accounting for optical depth effects.

We use this grid to calculate the chemical compositions and line luminosities by computing values of $n_{\rm H}$, $N_{\rm H}$, and $\alpha_{\rm vir}$ for each grid cell, and then performing a trilinear interpolation on the table. We calculate the number density of H nuclei as 
\begin{equation}
    n_{\mathrm{H}} = \frac{\rho}{m_{\mathrm{H}}},
\end{equation}
where $\rho$ is the total gas density of the cell and $m_{\rm H} = 2.34\times 10^{-24}$~g is the mass per H nucleus for Milky Way abundances. As for the column density of H nuclei $N_{\mathrm{H}}$, galactic-disc-scale simulations with ray-tracing-based radiative transfer and chemical network integration have shown that local models where photoshielding is accounted for with locally computed prescriptions perform reasonably well at reproducing the distribution and amount of molecular gas as compared with a detailed, global ray-tracing calculation \citep{SafranekShraderEtAl2017}. The local model is that the column density of H nucleus is defined as 
\begin{equation}
    N_{\mathrm{H}} = n_{\mathrm{H}} L_{\mathrm{shield}},
\end{equation}
where $L_{\mathrm{shield}}$ is the shielding length. \citet{SafranekShraderEtAl2017} show that an approach based on the Jeans length ($L_{\mathrm{shield}} = L_{\mathrm{J}} = (\pi c_{\mathrm{s}}^2 / \mathrm{G} \rho)^{1/2}$ where $c_{\mathrm{s}}$ is the local sound speed), with a $T = 40$ K temperature cap on the temperature used to compute the sound speed, yields the most accurate H$_2$ and CO abundances out of all the local approximations tested. Finally, we estimate the cell-by-cell virial parameter as
\begin{equation}
    \alpha_{\mathrm{vir}} = \frac{20}{3 \pi \mathrm{G}} \frac{\sigma^2 m_{\mathrm{H}} n_{\mathrm{H}}}{(m_{\mathrm{H}} N_{\mathrm{H}})^2},
\end{equation}
where $\sigma$ is the velocity dispersion from the mass-weighted mean gas velocity of the cell averaged over the surrounding 27 cells. (The equivalence between this form of the virial parameter and the more common \citet{BertoldiMcKee1992} form, phrased in terms of the mass and radius of a cloud, is straightforward to demonstrate algebraically.)

For the observational comparisons that we will carry out below, we require synthetic maps of tracers of star formation in addition to the atomic and molecular line luminosities. We produce synthetic H$\alpha$ maps directly from our simulations as follows. For every star particle formed in our simulations, we know the instantaneous ionising luminosity from \textsc{slug}, and we convert this to an H$\alpha$ luminosity using $L_{\rm H\alpha} = 1.0\times 10^{-12}$~erg~photon$^{-1}$ \citep{daSilvaEtAl2014}. This conversion assumes that 73\% of the ionising photons injected are eventually absorbed by hydrogen, while the remaining 27\% are absorbed by dust grains \citep{WilliamsMcKee1997}. We assign this H$\alpha$ luminosity to the cell that hosts each star particle. We do not include the effects of extinction, since real measurements of star formation rates in Milky Way-like galaxies are usually extinction corrected using the IR luminosity, the Balmer decrement, or some other method \citep{KennicuttEvans2012}.

\section{Results}
\label{Results}

Here we describe the results of our simulation and how it compares to observed galaxies, proceeding from large to small scales. First, in \autoref{Morphology of the galactic disc}, we show the morphology of the galactic disc and mock observations of molecular gas and star forming regions. In \autoref{Integrated constraints}, we show comparisons between our simulation outputs and integrated properties of the Milky Way and similar galaxies: the time evolution of the total star formation rate and the masses of various ISM components (cold neutral medium, warm neutral medium, and molecular gas). In \autoref{Large-scale constraints: radial profiles}, we show the large-to-medium scale constraints: radial profiles of surface density, velocity dispersion, and Toomre $Q$ parameter. In \autoref{Meso-scale constraints: spatially-resolved star formation relation}, we show the most important meso-scale constraint, the spatially-resolved star formation relation. In \autoref{Small-scale Constraints: GMCs}, we show the small-scale constraints: the distributions of GMC properties, scaling relations, and GMC lifetimes. In those sections, we show that our simulation satisfy all of these observational constraints. However, in \autoref{Cloud-scale Constraints: SF Uncertainty Principle} we show that our simulation fails to reproduce the observed scale-dependence of the correlation between star formation and molecular gas tracers.

\subsection{Morphology of the galactic disc}
\label{Morphology of the galactic disc}

\begin{figure*}
\includegraphics[width=\hsize]{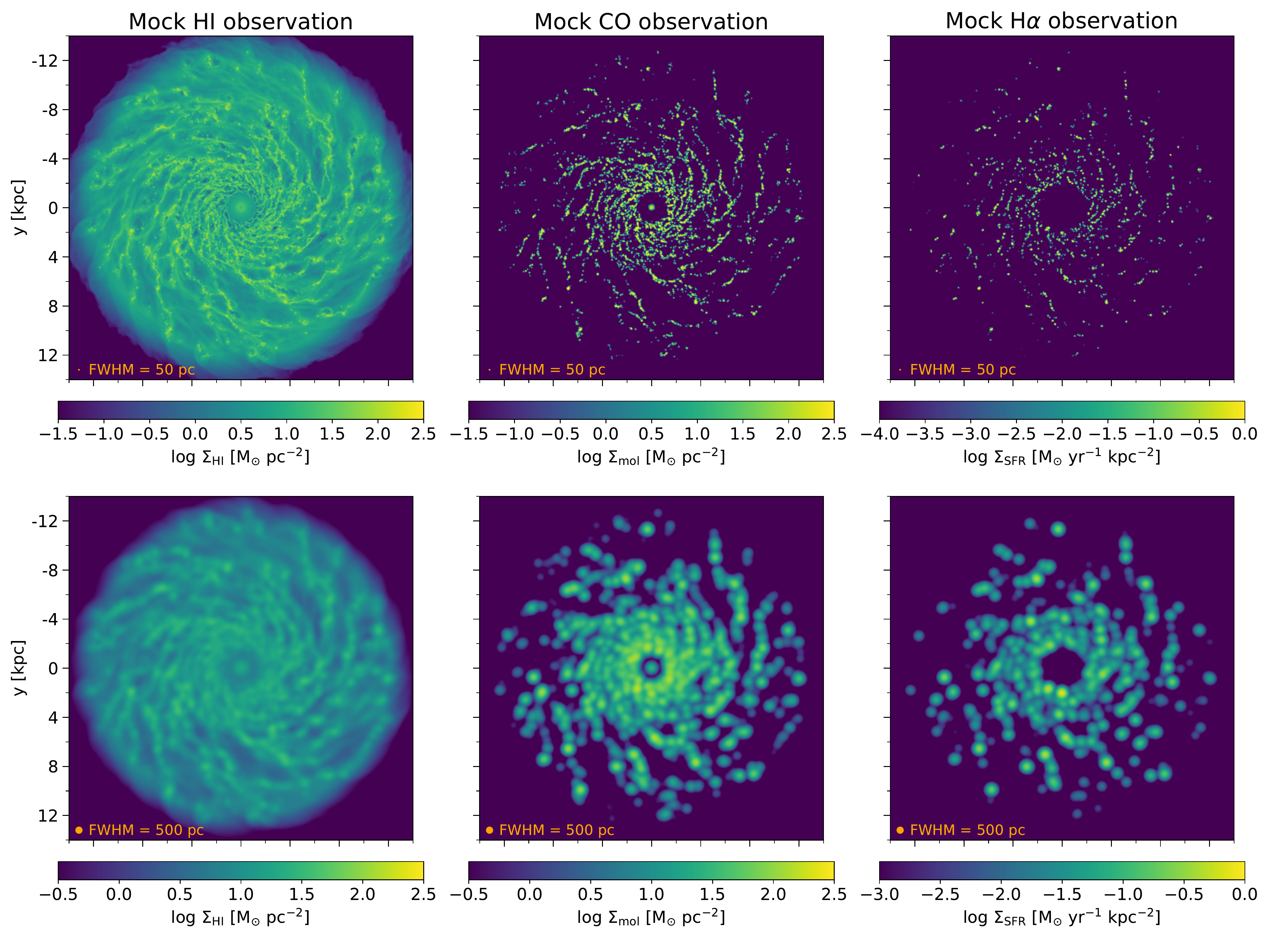}
\caption{
Morphology of the simulated galactic gas disc and mock observations of the H\textsc{i}-21 cm (left column), CO $J = 1 \rightarrow 0$ (middle column) and H$\alpha$ (right column) emission lines. Mock observations have been convolved with two-dimensional Gaussian kernels with standard deviation $\sigma = 21.2$ pc ($\mbox{FWHM} = 50$ pc; \textit{top row}) and 212 pc ($\mbox{FWHM} = 500$ pc; \textit{bottom row}); the beam sizes are shown as orange circles at the bottom left corner of each panel. Rather than showing the raw line luminosities, we have converted them to gas masses and star formation rates using the same methods commonly-adopted for observations. For H\textsc{i} we assume the gas is optically thin, so the transformation between mass and line luminosity is trivial. We convert CO luminosity to mass as $M_{\mathrm{mol}} = \alpha_{\mathrm{CO}} L_{\mathrm{CO}}$, where $\alpha_{\mathrm{CO}} = 4.3\ \mathrm{M_{\odot}} (\mathrm{K\ km\ s^{-1}\ pc^{2}})^{-1}$ \citep{BolattoWolfireLeroy2013}, and H$\alpha$ luminosity to star formation rate as $\mbox{SFR} = \alpha_{\rm SF} \alpha_{\rm H\alpha} L_{\rm H_\alpha}$, where $\alpha_{\rm SF} = 1.08 \times 10^{-53}\ \mathrm{M_{\odot}}\ \mathrm{yr}^{-1} (\mathrm{photons\ s^{-1}})^{-1}$ is the conversion between ionising luminosity and star formation rate \citep{KennicuttTamblynCongdon1994, MadauPozzettiDickinson1998, Kennicutt1998} and  $\alpha_{\mathrm{H}\alpha} \approx 1.0\times 10^{12}$ photon erg$^{-1}$ is the number of ionising photons required per unit energy radiated into the H$\alpha$  \citep{daSilvaEtAl2014}.
}
\label{fig:projections}
\end{figure*}

\autoref{fig:projections} shows mock observations of our galaxy in the H\textsc{i}-21 cm (left), CO $J = 1 \rightarrow 0$ (middle) and H$\alpha$ (right) emission lines. To mimic the finite resolution of observations, we show maps convolved with two-dimensional Gaussian kernels with standard deviations of $\sigma = $ 21.2 pc and 212 pc, corresponding to FWHM sizes $l = $ 50 pc (top row) and 500 pc (bottom row). Note that the initial gas distributions in the galaxy centre of $r < 2$ kpc and the outer-disc region of $r > 14$ kpc are set as non-realistic low density gas, and that star formation is not allowed to occur in those regions. Therefore, we see a very smooth gas distribution without any fragmentation of the gas in the galactic centre. In this paper, we will focus on the main disc regions between radii of $r = 3 - 11$ kpc.

The overall morphology in H\textsc{i}, CO, and H$\alpha$ is very similar to that observed in nearby flocculent spiral galaxies at similar resolution (e.g.,~NGC 628 or NGC 4254 -- c.f.~\citealt{Walter08a} for H\textsc{i}, \citealt{Sun18a} for CO, \citealt{Kreckel18a} for H$\alpha$). The most obvious morphological difference is that the H$\alpha$ in \autoref{fig:projections} is more similar to the CO than appears to be the case in, for example, NGC 628. This is a significant point, to which we shall return below.

\subsection{Integrated constraints}
\label{Integrated constraints}

\subsubsection{Star formation rate}
\label{star formation rate}

\begin{figure}
\includegraphics[width=\columnwidth]{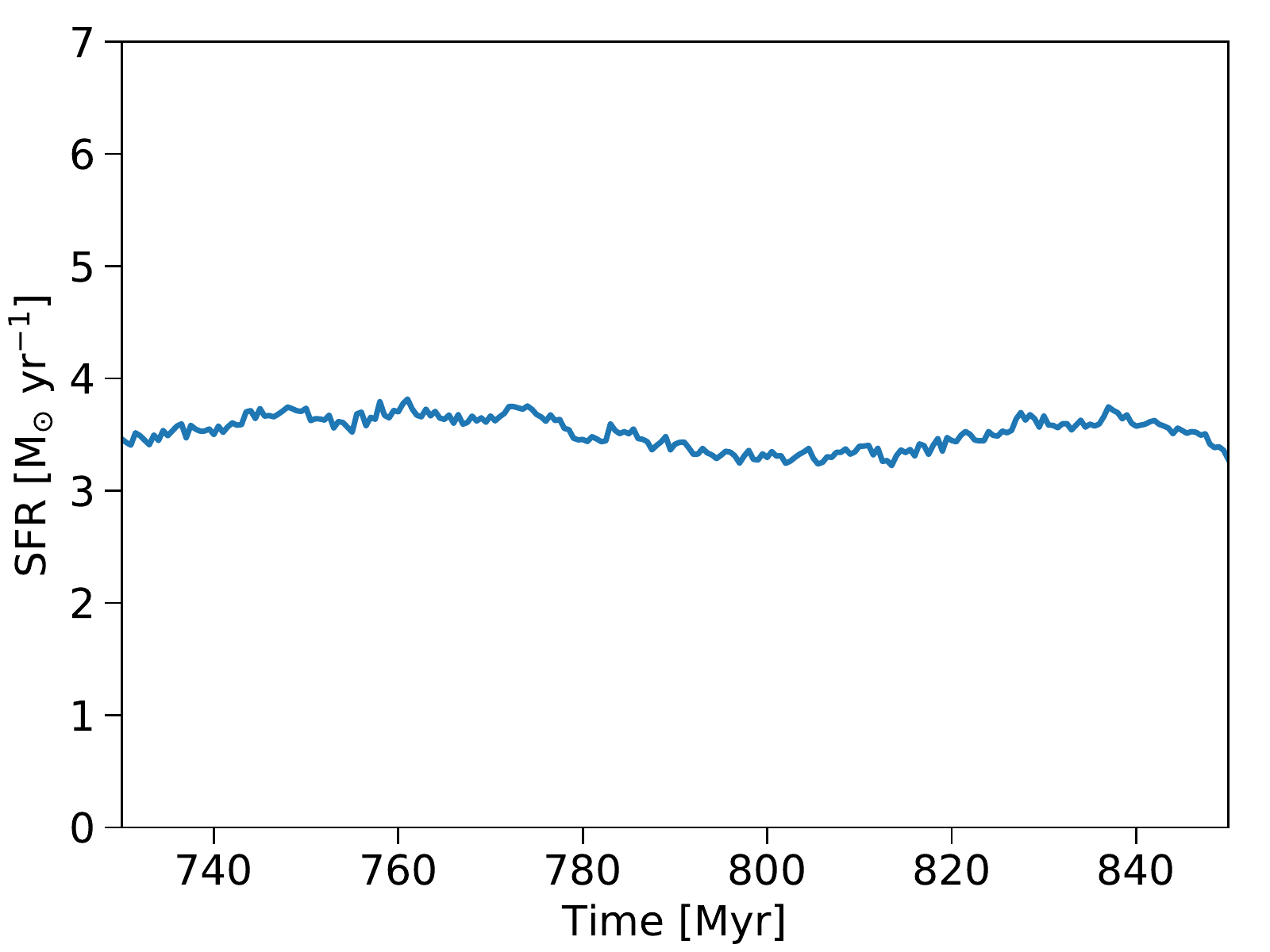}
\caption{Time evolution of SFR. The SFR is measured by counting newly formed star particles with ages $\leq$ 1 Myr rather than using our simulated H$\alpha$ emission, but by construction these two methods would give nearly identical results for the integrated SFR of the entire galaxy.
\label{fig:time_SFR}}
\end{figure}

The first observational constraint to which we compare is the total star formation rate (SFR) in the galaxy, which we show as a function of time in \autoref{fig:time_SFR}. Note that we show the total SFR in the whole disc, not only in the main disc between radii of $r = $ 3 - 11 kpc. The SFR does not change drastically with time, showing that the galactic disc has reached a quasi-equilibrium state. In the equilibrium state, the SFR is $\sim 3.4$ $\mathrm{M_{\odot}\ {yr}^{-1}}$, which is within a factor of $\approx 2$ of the observed Milky Way SFR of $1 - 2\ \mathrm{M_{\odot}\ yr^{-1}}$ \citep{RobitailleWhitney2010, MurrayRahman2010, ChomiukPovich2011, LicquiaNewman2015}. Considering the uncertainty of the Galactic properties such as the total gas mass and the scale radius/height that inform the initial conditions of the galaxy simulation, this is well within the error bar. In this connection, other Milky Way-type simulations which use similar stellar feedback models to ours also show SFRs of $2 - 5\ \mathrm{M_{\odot}\ yr^{-1}}$ \citep{HopkinsQuataertMurray2012, GoldbaumKrumholzForbes2016, GrisdaleEtAl2017}. This SFR places our simulated galaxy squarely on the integrated Kennicutt-Schmidt relation \citep[e.g.,][]{KennicuttEvans2012}.

\subsubsection{ISM phase structure}
\label{ISM phase structure}

\begin{figure}
\includegraphics[width=\columnwidth]{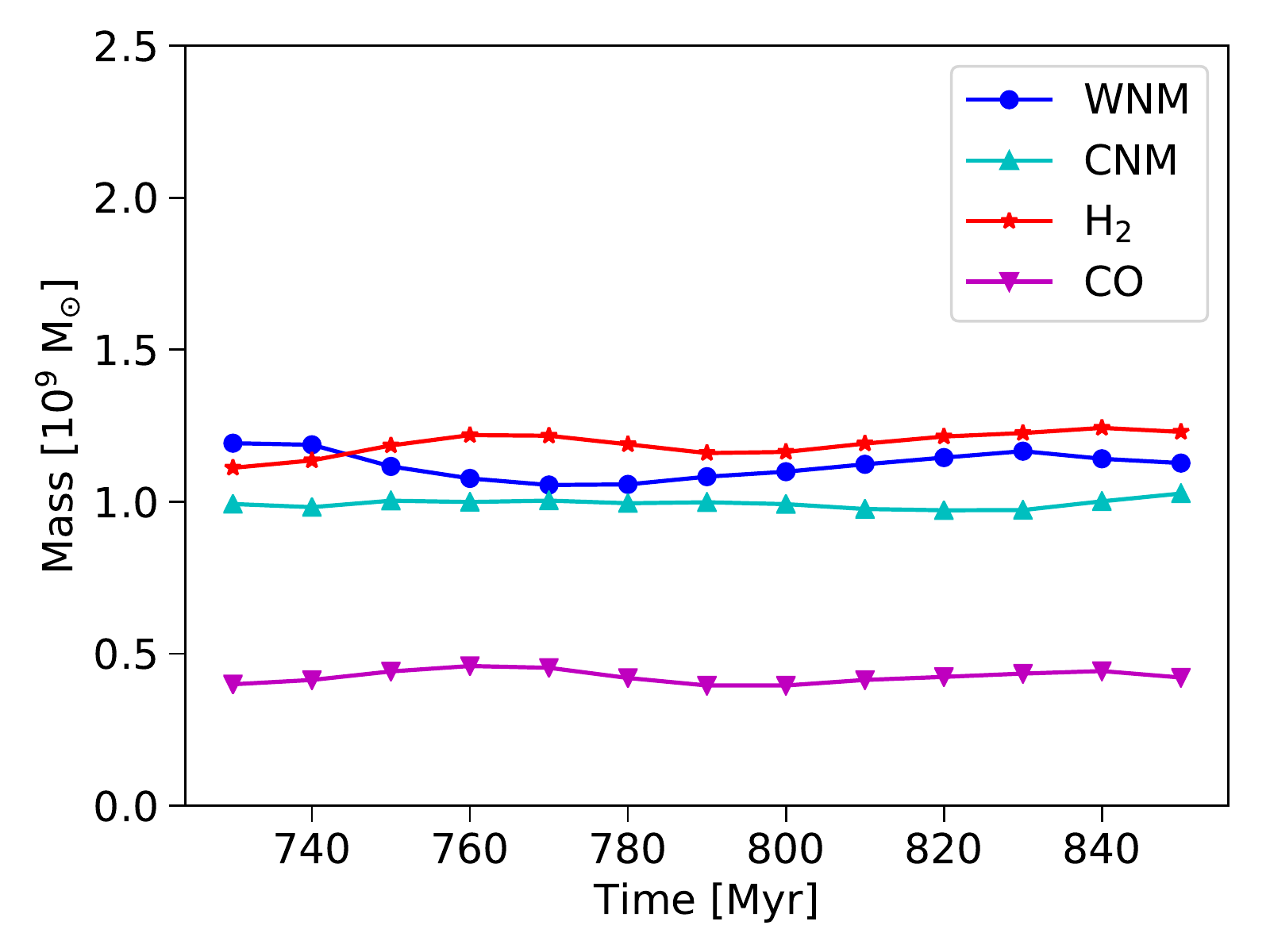}
\caption{Time evolution of the masses of various ISM components: warm neutral medium (WNM), cold neutral medium (CNM), $\mathrm{H_2}$ and molecular gas traced by CO. See the main text for an explanation of how we define each of these phases.}
\label{fig:ISM_components}
\end{figure}

\autoref{fig:ISM_components} shows the time evolution of the masses of several ISM components: warm neutral medium (WNM), cold neutral medium (CNM), $\mathrm{H_2}$ and molecular gas traced by CO. The WNM is defined as H\textsc{i} gas whose temperature $T$ is between $5 \times 10^3$ K and $10^4$ K and with number density higher than $10^{-2}\ \mathrm{cm}^{-3}$. The CNM is defined as $\mathrm H\textsc{i}$ gas whose temperature is lower than $10^3$ K. The H$_2$ mass, $M_{\rm H_2}$, shown in the figure is the mass of gas that is chemically H$_2$, as distinct from the molecular gas mass traced by CO, $M_{\rm mol}$, which denotes the mass of gas within which carbon is chemically in the form of CO; formally, we define the $M_{\rm mol} = f_{\rm CO} M_{\rm cell} = (n_{\rm CO}/n_{\rm C}) M_{\rm cell}$, where the sum runs over all cells, $M_{\rm cell}$ is the cell mass, and $n_{\rm CO}$ and $n_{\rm C}$ are the numbers of CO molecules and C nuclei per H atom, respectively. 

As with the total SFR, we see no time-dependence in the ISM phase structure, showing again that the galactic disc is in a quasi-equilibrium state. We find that the bulk of the ISM is in the WNM or CNM rather than the molecular phase. In observations, the typical molecular-to-atomic hydrogen mass ratio in Milky Way-sized galaxies is $M_{\rm mol} /M_{\rm HI} \sim 0.3$ with significant scatter of $\pm 0.4$ dex \citep{SaintongeEtAl2011}; by comparison, our simulations give $M_{\rm mol}/(M_{\rm WNM} + M_{\rm CNM}) \approx 0.2$, where we use $M_{\rm mol}$ rather than $M_{\rm H_2}$ to be consistent with observation that use CO-based molecular masses. If we instead use $M_{\rm H_2}$, we obtain $\sim$ 0.5. Given the large scatter in observed molecular-to-atomic ratios, either result is consistent with the observations.

The mass ratio of the molecular gas traced by CO over total $\mathrm{H_2}$ is $M_{\rm mol}/M_{\rm H_2} \approx 0.4$, so slightly more than half the molecular gas is in the form of CO-dark H$_2$. This is consistent with estimates that 30 - 70\% of the Milky Way's molecular gas is CO-dark based on gamma-ray observations \citep[e.g.,][]{GrenierCasandjianTerrier2005}, dust continuum emission \citep[e.g.,][]{PlanckCollaboration2011} and $\mathrm C^{+}$ 158 $\mathrm{\mu m}$ emission \citep[e.g.,][]{PinedaEtAl2013, LangerEtAl2014}. It is also in line with estimates from previous analytic models and cosmological zoom simulations \citep{WolfireHollenbachMcKee2010, LiEtAl2018b}.

\subsection{Large-scale constraints: radial profiles}
\label{Large-scale constraints: radial profiles}

\begin{figure*}
\includegraphics[width=160pt]{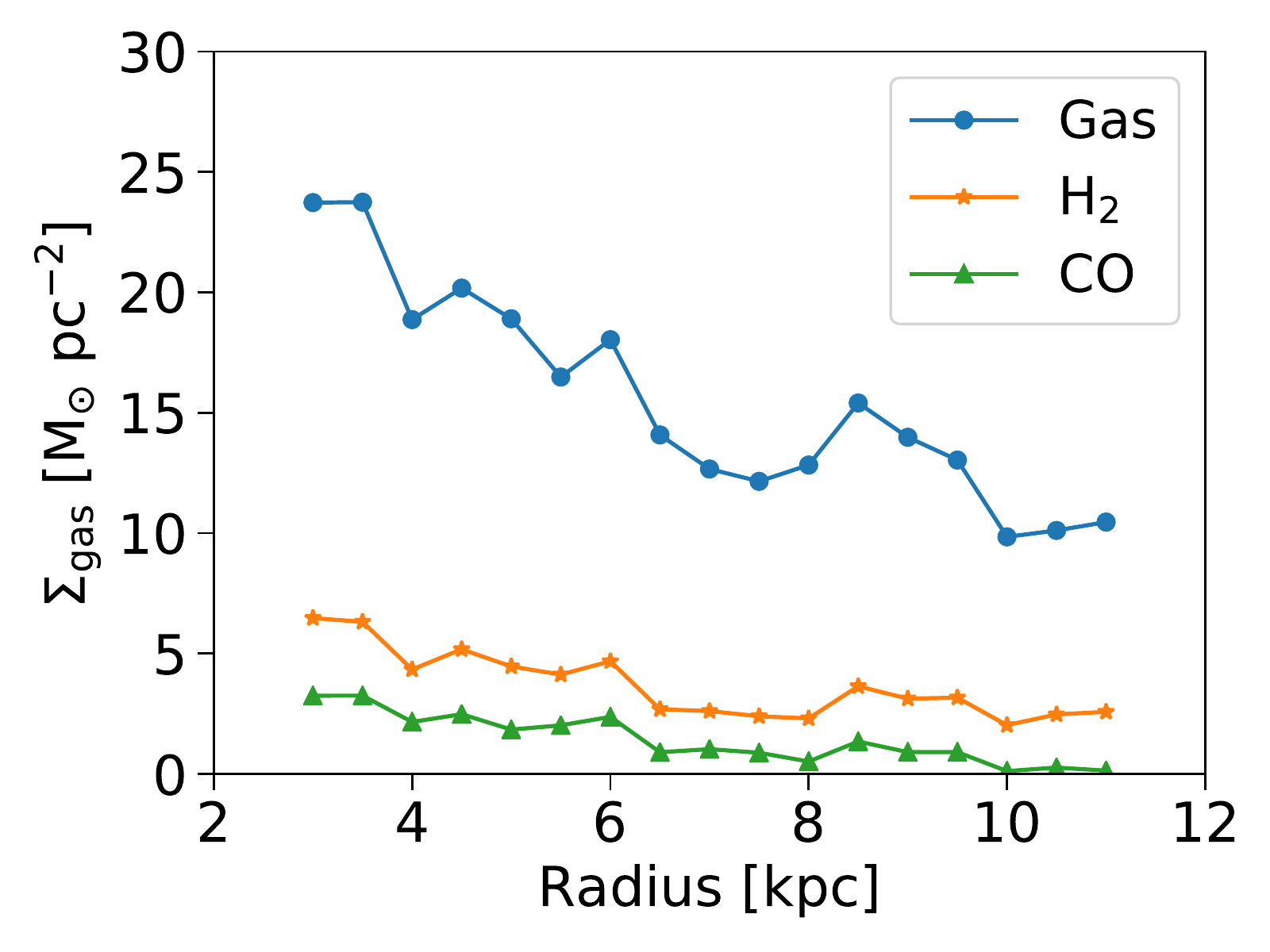}
\includegraphics[width=160pt]{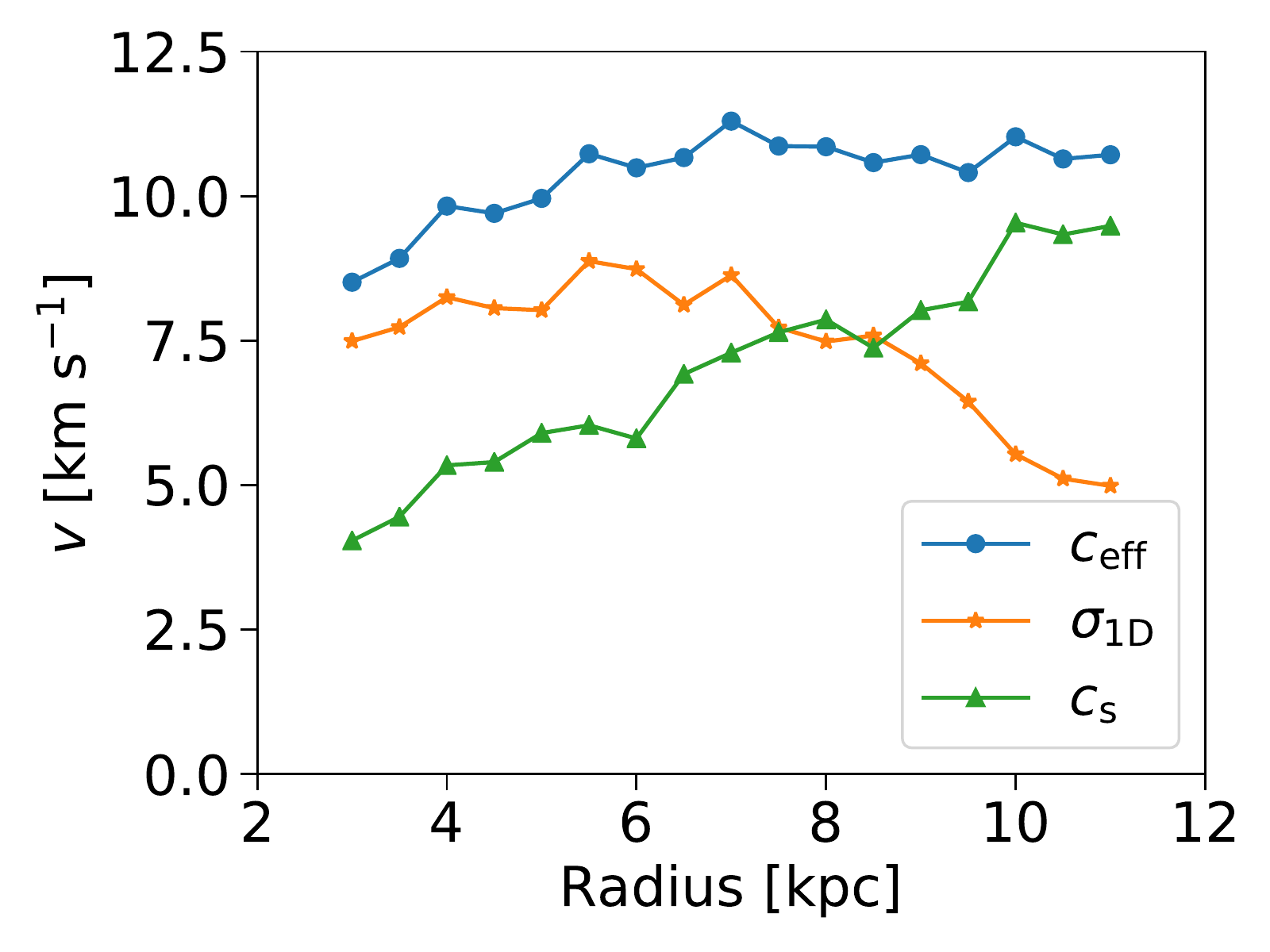}
\includegraphics[width=160pt]{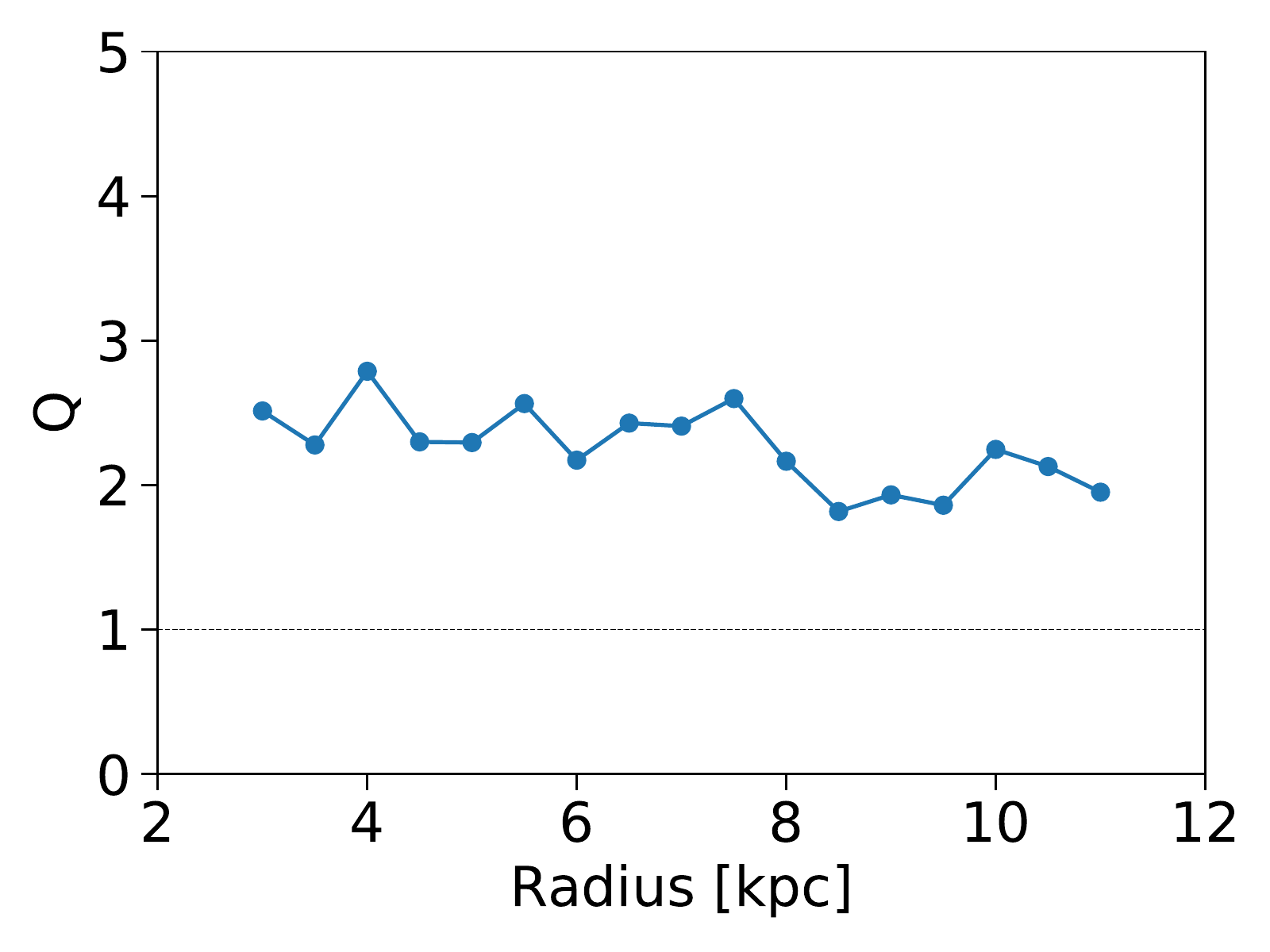}
\caption{Azimuthally-averaged radial profiles for the galactic disc. The panel shows surface density of total gas, $\mathrm{H_2}$, and molecular gas traced by CO (\textit{left}), mass-averaged one-dimensional non-thermal velocity dispersion $\sigma_{\rm 1D}$, sound speed $c_s$, and total velocity dispersion $c_{\rm eff} = (\sigma_{\rm 1D}^2+c_s^2)^{1/2}$ (\textit{middle}), and Toomre $Q$ parameter (\textit{right}). 
}
\label{fig:radial_profiles}
\end{figure*}

We next zoom in to slightly smaller scales, and examine the radial profiles of various gas properties in our galaxy, with attention to the kinematics and phase structure as a function of radius. The left panel of \autoref{fig:radial_profiles} shows the gas surface density azimuthally-averaged over 500 pc-wide radial bins. This is an input rather than an output for our simulations, but we include it to show that we have a relatively flat distribution of gas with radius, with a surface density of $10 - 20\ \mathrm{M_{\odot}\ pc^{-2}}$, consistent with the Milky Way's value of $\sim 10\ \mathrm{M_{\odot}\ pc^{-2}}$ \citep[e.g.][]{WolfireEtAl2003, YinEtAl2009}.
The surface density of $\mathrm{H_2}$ mass averaged over these large scales is much below 10 $\mathrm{M_{\odot}\ pc^{-2}}$, consistent with values seen in Milky Way-like nearby galaxies \citep{BigielEtAl2008}. As we already discussed in the context of \autoref{fig:ISM_components}, we again see that slightly more than half the H$_2$ is CO-dark, and that this fraction does not vary strongly with radius. 

The middle panel of \autoref{fig:radial_profiles} shows the mass-weighted mean non-thermal gas velocity dispersion as a function of radius. We define this quantity as 
\begin{equation}
\sigma_{\mathrm{1D}} = \sqrt{[\mathbfit{v} - \mathbfit{v}_{\mathrm{circ}}(r)]^2/3},
\end{equation}
where $\mathbfit{v}_{\mathrm{circ}}(r)$ is the mass-weighted mean gas circular velocity. The weighted mean is computed over all the gas in a given radial bin that lies within $\pm 1$ kpc of the disc midplane. We also compute the mass-weighted mean thermal sound speed, 
\begin{equation}
c_{\mathrm{s}} = \sqrt{\gamma(\gamma - 1)e},
\end{equation}
where $\gamma = 5/3$ is the adiabatic index and $e$ is the thermal energy density, and the total velocity dispersion, 
\begin{equation}
c_{\mathrm{eff}} = \sqrt{\sigma_{\mathrm{1D}}^2 + c_{\mathrm{s}}^2}. 
\end{equation}
This final quantity is the total velocity dispersion that would be recovered in a spectroscopically resolved observation of emission lines. One important observational constraint that we expect our simulations to match is that all nearby galaxies have nearly constant velocity dispersions of $\approx 10\ \mathrm{km\ s^{-1}}$ at all radii across their discs \citep[excluding the inner few hundred pc; e.g.,][]{TamburroEtAl2009, IanjamasimananaEtAl2012}, with no significant difference in velocity dispersion between H\textsc{i} and CO \citep{Caldu-Primo13a}. \autoref{fig:radial_profiles} shows that our simulations do an excellent job of reproducing this observation. 

The right panel of \autoref{fig:radial_profiles} shows the Toomre $Q$ parameter, 
\begin{equation}
Q = \frac{\kappa c_{\mathrm{eff}}}{\pi G \Sigma_{\mathrm{gas}}}, 
\end{equation}
where $\kappa$ is the epicycle frequency defined by 
\begin{equation}
\kappa = \sqrt{2} \frac{v_{\mathrm{circ}}}{r} \left(1 + \frac{r}{v_{\mathrm{circ}}} \frac{dv_{\mathrm{circ}}}{dr} \right)^{1/2}.
\end{equation}
At all radii, the disc has $Q \sim 2$, indicating that the galactic disc is marginally gravitationally stable. This is again consistent with nearby spiral galaxies \citep{LeroyEtAl2008}.

\subsection{Meso-scale constraints: spatially-resolved star formation relation}
\label{Meso-scale constraints: spatially-resolved star formation relation}

\begin{figure*}
\includegraphics[width=\hsize]{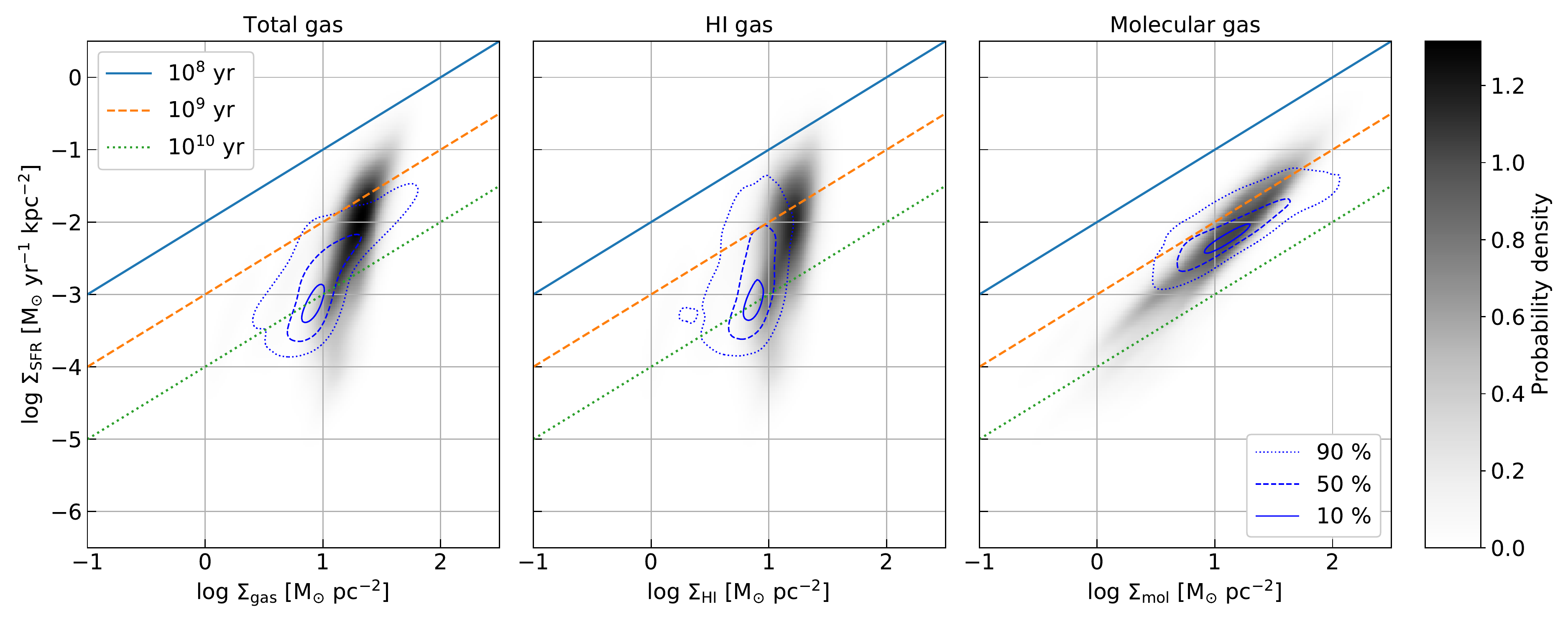}
\caption{SFR surface density as a function of total gas (left), H \textsc{i} (middle) and molecular gas (right) surface density. Grayscale shows the probability density for pixels in our simulation. All quantities shown are those derived from simulation observations rather than the true simulation values -- see \autoref{fig:projections}. We degrade the resolution of our surface density maps to 750 pc to derive the data points in this plot. We also mark lines of constant depletion time ($10^8$, $10^9$ and $10^{10}$ years) as solid, dashed and dotted lines. The blue contours show the nearby galaxy observations of \citet{BigielEtAl2008} and mark the loci that encompass 90 per cent (dotted), 50 per cent (dashed), and 10 per cent (solid) of the data.}
\label{fig:KS_plot}
\end{figure*}

Perhaps one of the most important constraints on star formation in resolved galaxy simulations is the relationship between H\textsc{i}, H$_2$, and star formation measured at kpc scales.
\autoref{fig:KS_plot} shows the SFR surface density as a function of the surface density of the total gas (left), $\mathrm H\textsc{i}$ gas (middle) and molecular gas (right). The SFR is measured by evaluating the star formation formula applied in the simulation; we do not use the H$\alpha$ luminosity. To obtain the $\mathrm H\textsc{i}$ and molecular gas mass, we use the H\textsc{i}-21 cm and CO $J = 1 \rightarrow 0$ line emission converted to these quantities, rather than using the true masses, as in \autoref{fig:projections}. To compare with observations, which are typically carried out for resolution elements of hundreds of pc, we degrade the resolution of our surface density maps to 750 pc, matching the resolution of the THINGS survey \citep{BigielEtAl2008, LeroyEtAl2008}. 

We see that our simulations again provide a reasonable quantitative match to observations. The relationship between total gas surface density and star formation rate is superlinear, driven largely by the fact that the H\textsc{i} surface density shows a hard maximum at $\approx 10$ $M_\odot$ pc$^{-2}$, so that $\Sigma_{\rm SFR}$ and $\Sigma_{\rm HI}$ are nearly-uncorrelated.\footnote{Because our simulations feature a galactic disc that is truncated at large radii, we do not include the region where total gas surface density $\Sigma_{\rm gas} \lesssim 1$ $M_\odot$ pc$^{-2}$, where H\textsc{i} and star formation are correlated -- \citet{Bigiel10a}.} On the other hand, the slope of the relationship between the surface densities of star formation and molecular gas is much shallower, similar to the roughly linear scaling between these two quantities seen in nearby galaxies \citep{BigielEtAl2008, LeroyEtAl2013}.

The simulation does not match the mean of the observations perfectly. For instance, the maximum $\mathrm H\textsc{i}$ gas surface density in the simulation is slightly higher than the observed value, and the slope in the molecular gas is slightly steeper than unity. However, the simulation results are well within the level of galaxy-to-galaxy variation that is actually seen among nearby galaxies \citep[e.g.~figs.~4 to~6 of][]{BigielEtAl2008}. Moreover, the results for the relation between total gas and star formation rate are very similar to those seen in galaxy simulations whose stellar feedback recipes are similar to ours \citep[e.g.][]{AgertzKravtsovLeitner2013, GoldbaumKrumholzForbes2016}, while those for the individual H\textsc{i} and H$_2$ components are slightly sensitive to our assumed cosmic ray ionisation rate and ISRF strength. Our chemical tables use constant values for these parameters, but in reality both should vary at least slightly with position in the galaxy. A more realistic treatment of this variation could easily alter the results at a level comparable to the small differences we see between our simulations and observations. 
Similarly, it is likely that our chemical results are at least somewhat sensitive to our resolution. Recent studies in simulations using on-the-fly chemical networks (as opposed to our post-processing) have found that a spatial resolution $\leq 0.1$ pc is required for H$_2$ and CO fractions to converge fully \citep{SeifriedEtAl2017}. The resolution requirements for our post-processing technique are likely somewhat less severe, since the need for high spatial resolution is driven largely by the need to capture short-timescale fluctuations in density and shielding, which we are averaging out. None the less, it is likely that changing the resolution would at least somewhat alter our chemical abundances.

\subsection{Small-scale Constraints: GMCs}
\label{Small-scale Constraints: GMCs}

\begin{figure*}
\includegraphics[width=160pt]{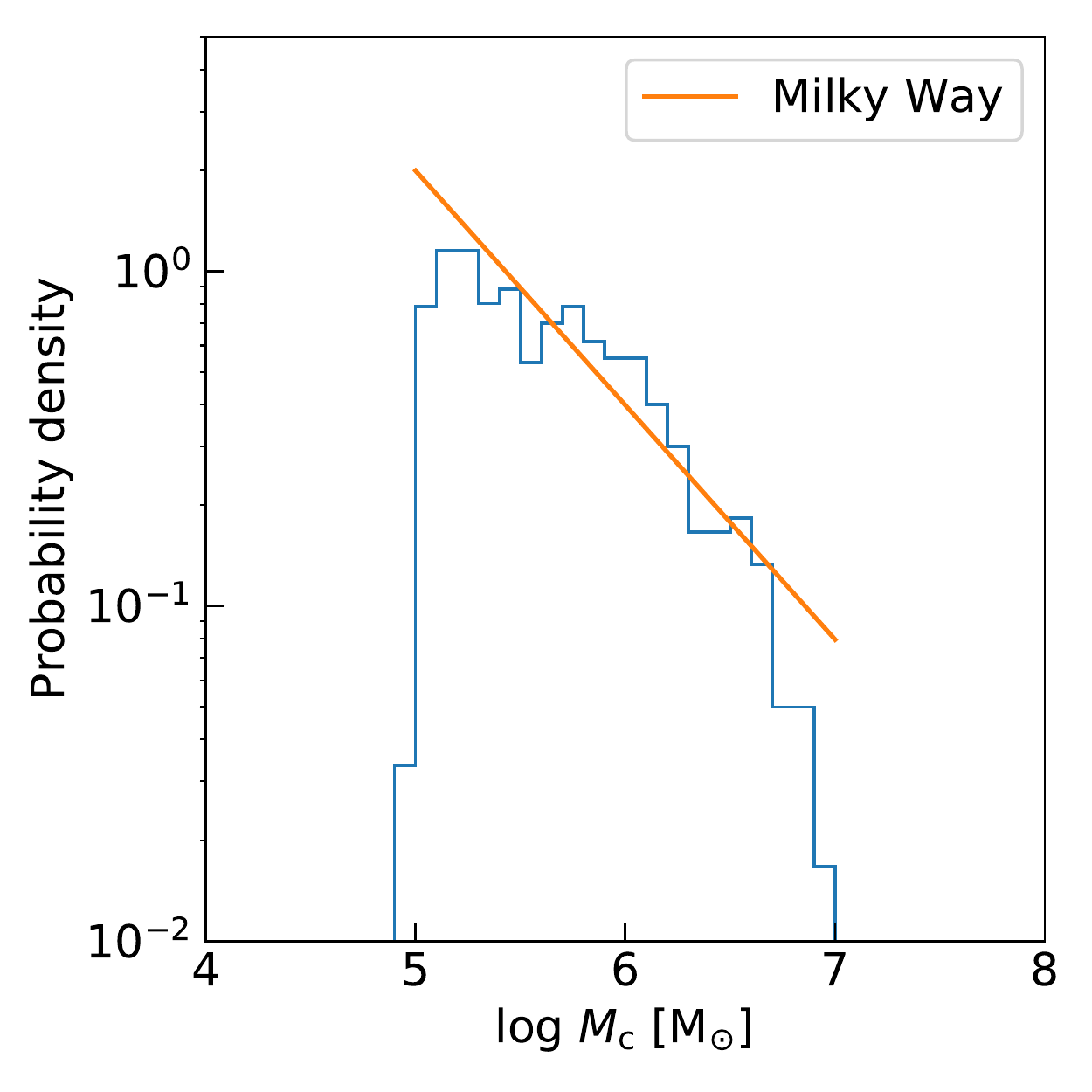}
\includegraphics[width=160pt]{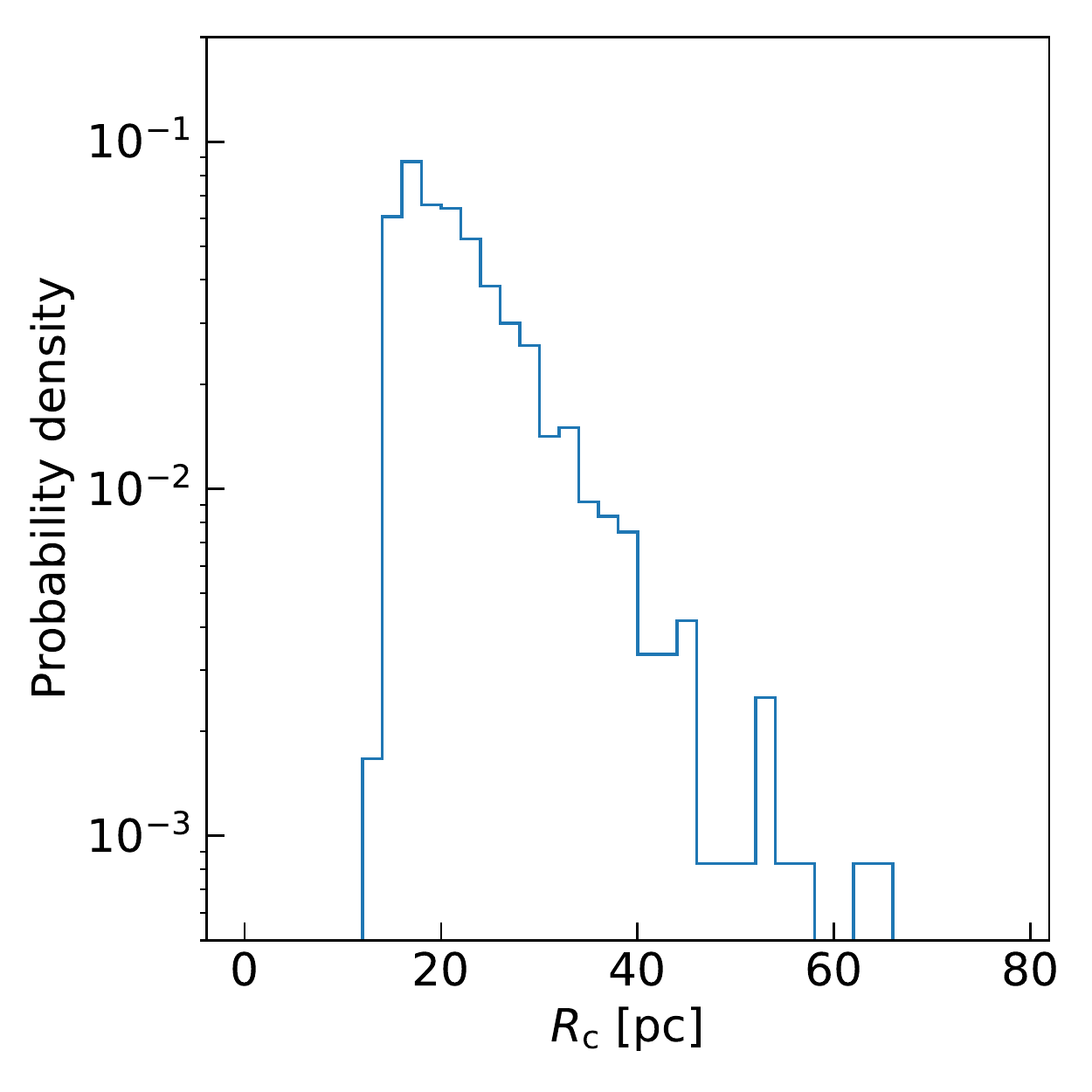}
\includegraphics[width=160pt]{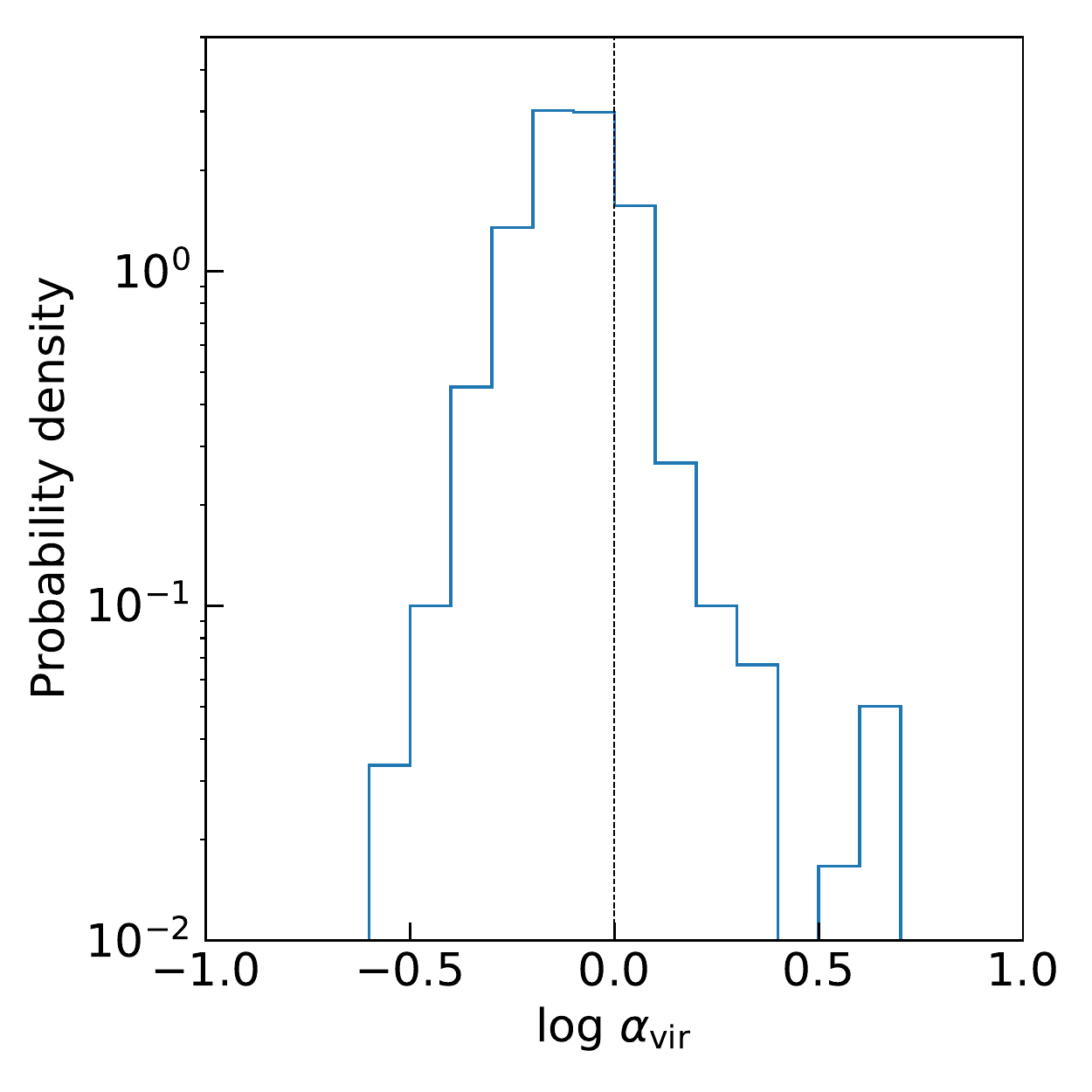}
\caption{Distribution of GMC masses (left), radii (middle), and virial parameters (right). Clouds are identified, and their properties are derived, using the procedure outlined in the main text; at the time shown in this figure (790 Myr), there are 600 identified GMCs. In the left panel, the orange line shows the observed Milky Way GMC mass spectrum, $dN/dM\propto M^{-\beta}$ with $\beta\approx 1.7$.}
\label{fig:cloud_properties}
\end{figure*}

We next examine the properties of GMCs produced in our simulations, which we define as connected structures within which $f_{\mathrm{CO}} \geq 0.3$; we exclude from consideration any such structures that contain fewer than $3^3$ cells, on the grounds that we cannot derive meaningful properties for such unresolved objects \citep{TaskerTan2009, FujimotoEtAl2014}. The left panel of \autoref{fig:cloud_properties} shows the number weighted probability density function of the cloud mass. There is a low mass cutoff around $M_{\mathrm{c}} \sim 10^5\ \mathrm{M_{\odot}}$ because we impose the threshold of the number of cells for cloud identification. The cloud mass ranges from $10^5\ \mathrm{M_{\odot}} < M_{\mathrm{c}} < 10^7\ \mathrm{M_{\odot}}$, in reasonable agreement with the GMCs observed in nearby galaxies \citep[e.g.][]{RosolowskyEtAl2003,FreemanEtAl2017}. We also compare the cloud mass spectrum to a power law distribution $dN / dM \propto M^{-\beta}$. For GMCs in the Milky Way, $\beta$ is consistently found to be in the range 1.6 to 1.8 over mass range $M_{\mathrm{c}} > 10^4\ \mathrm{M_{\odot}}$ \citep{SolomonEtAl1987, WilliamsMcKee1997, HeyerEtAl2009, RomanDuvalEtAl2010}. The slope of our cloud mass distribution is consistent with $\beta \sim 1.7$.

The middle panel of \autoref{fig:cloud_properties} shows the number weighted probability density function of cloud radius. We define the cloud radius as
\begin{equation}
    R_{\mathrm{c}} = \sqrt{\frac{A_{xy} + A_{yz} + A_{zx}} {3 \pi}},
\end{equation}
where $A_{xy}$ is the projected area of the cloud in the $x$-$y$ plane, $A_{yz}$ is that in the $y$-$z$ plane and $A_{zx}$ is that in the $z$-$x$ plane. The peak of the PDF of radius is near 20 pc and the largest GMCs have radii of $\sim$ 60 pc. There is a low radius cutoff around $R_{\mathrm{c}} < 20$ pc because we impose the threshold of the number of cells for cloud identification. The sizes of the clouds are comparable to those observed for GMCs in the Galaxy \citep[e.g.][]{MivilleDeschenesMurrayLee2017}.

The right panel of \autoref{fig:cloud_properties} shows the number weighted probability density function of the cloud virial parameter defined as,
\begin{equation}
    \alpha_{\mathrm{vir}} = \frac{5\sigma^2_{\mathrm{c}} R_{\mathrm{c}}} {G M_{\mathrm{c}}} = \frac{5 (\sigma^2_{\mathrm{1D}} + c^2_{\mathrm{s}}) R_{\mathrm{c}}} {G M_{\mathrm{c}}},
\end{equation}
where $\sigma_{\mathrm{1D}}$ is the mass-averaged one-dimensional velocity dispersion defined as $\sigma_{\mathrm{1D}} = \langle\sqrt{(\mathbfit{v} - \mathbfit{v}_{\mathrm{CoM}})^2/ 3}\rangle$, where $\mathbfit{v}$ is the velocity of the gas and $\mathbfit{v}_{\mathrm{CoM}}$ is the cloud's centre of mass velocity, $c_{\mathrm{s}}$ is the sound speed, and the angle brackets indicate a mass-weighted average over the cells in the cloud \citep{BertoldiMcKee1992}. The distribution shows that the peak is around $\alpha_{\mathrm{vir}} \sim 1$, indicating that most clouds are gravitationally bound. This is consistent with observations. Virial parameters for clouds in the Milky Way and nearby galaxies exhibit a range of values from $\alpha_{\rm vir} \sim 0.1$ to $\alpha_{\rm vir} \sim 10$, but typically $\alpha_{\rm vir}$ is $\sim$ 1 \citep[e.g.][]{SolomonEtAl1987, BolattoEtAl2008, HeyerEtAl2009, RomanDuvalEtAl2010, WongEtAl2011}. 

\begin{figure}
\includegraphics[width=\columnwidth]{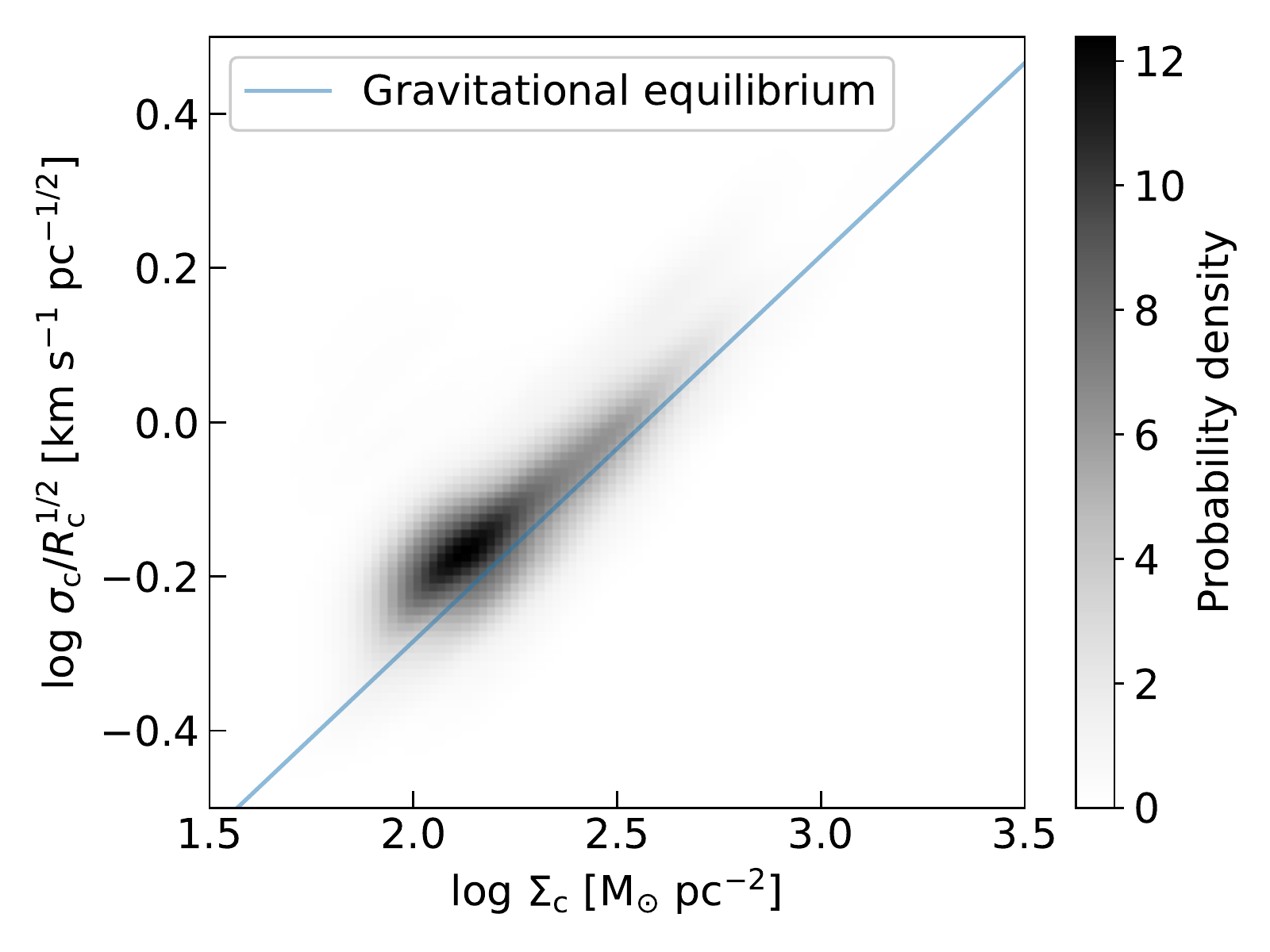}
\caption{Scaling relation of clouds between $\sigma_{\mathrm{c}}/R_{\mathrm{c}}^{1/2}$ and $\Sigma_{\mathrm{c}}$. The panel shows the joint probability density distribution. The solid line indicates gravitational equilibrium, i.e.~$\sigma_{\mathrm{c}}/R_{\mathrm{c}}^{1/2} = (\pi G / 5)^{1/2} \Sigma_{\mathrm{c}}^{1/2}$.}
\label{fig:cloud_scaling_relation}
\end{figure}

In \autoref{fig:cloud_scaling_relation}, we show the ``extended'' \citep{Larson1981} relation, which expresses the relationship between velocity dispersion $\sigma_{\mathrm{1D}}$, radius $R_{\mathrm{c}}$ and surface density $\Sigma_{\mathrm{c}}$ required for virial balance,
\begin{equation}
    \sigma_{\mathrm{c}}/R_{\mathrm{c}}^{1/2} = (\pi G / 5)^{1/2} \Sigma_{\mathrm{c}}^{1/2}.
\end{equation}
Our clouds follow this gravitational equilibrium line, again consistent with observations of GMCs in the Milky Way \citep{HeyerEtAl2009, RomanDuvalEtAl2010} and similar nearby galaxies \citep{Sun18a}.

\begin{figure}
\includegraphics[width=\columnwidth]{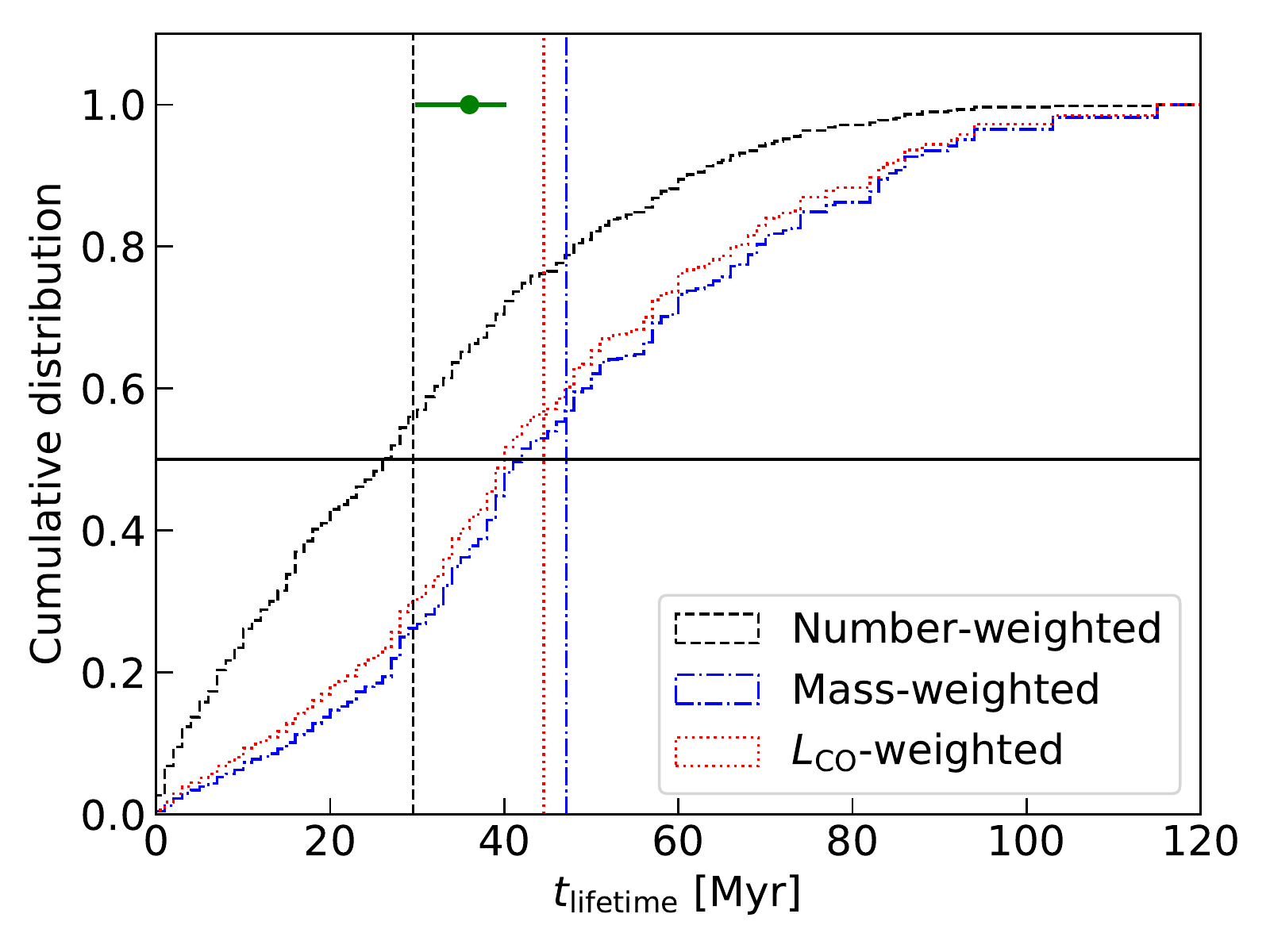}
\caption{Number-weighted (black dashed), mass-weighted (blue dash-dotted), and CO luminosity-weighted (red dotted) cumulative distribution function (bin size is 1.0 Myr) of the cloud lifetime at $t= $ 790 Myr, computed on our sample of 600 GMCs. The vertical lines show mean lifetimes; number-weighted (black dashed), mass-weighted (blue dash-dotted), and CO luminosity-weighted (red dotted). The horizontal line shows the probability of 0.5. The green point with error bars shows an average GMC lifetime of $36^{+4}_{-6}$ Myr obtained from the star formation `uncertainty principle' described in \autoref{Cloud-scale Constraints: SF Uncertainty Principle}.}
\label{fig:cloud_lifetime}
\end{figure}

Finally, we examine the distribution of GMC lifetimes in \autoref{fig:cloud_lifetime}. We determine this quantity by starting with the list of GMCs present at our fiducial time slice at 790 Myr. We then run the same GMC identification on snapshots at intervals of 0.5 Myr between $t= $ 730 and 850 Myr, with clouds in adjacent time outputs identified with one another following the method of \citet{TaskerTan2009} and \citet{FujimotoEtAl2014}. We use this method to compute the lifetimes of all clouds that are present at 790 Myr. Note that our method potentially underestimates the lifetimes of the longest-lived clouds, since our time baseline allows only a maximum 60 Myr lifetime before and 60 Myr lifetime after our central time of 790 Myr. However, we see that the mass-weighted median lifetime is only $\sim$ 40 Myr, so the effects of truncation are minimal and our median lifetime estimate is robust. While this quantity is difficult to determine from observation, there are a number of estimates available in the literature \citep[e.g.][]{KawamuraEtAl2009, MiuraEtAl2012, MeidtEtAl2015, kruijssen19, chevance19}. Our result is consistent with the observational estimates, which span lifetimes in the range 10--50~Myr, as well as with the theoretical model of \citet{jeffreson18}.

\subsection{Cloud-scale Constraints: SF Uncertainty Principle}
\label{Cloud-scale Constraints: SF Uncertainty Principle}

\begin{figure}
\includegraphics[width=\columnwidth]{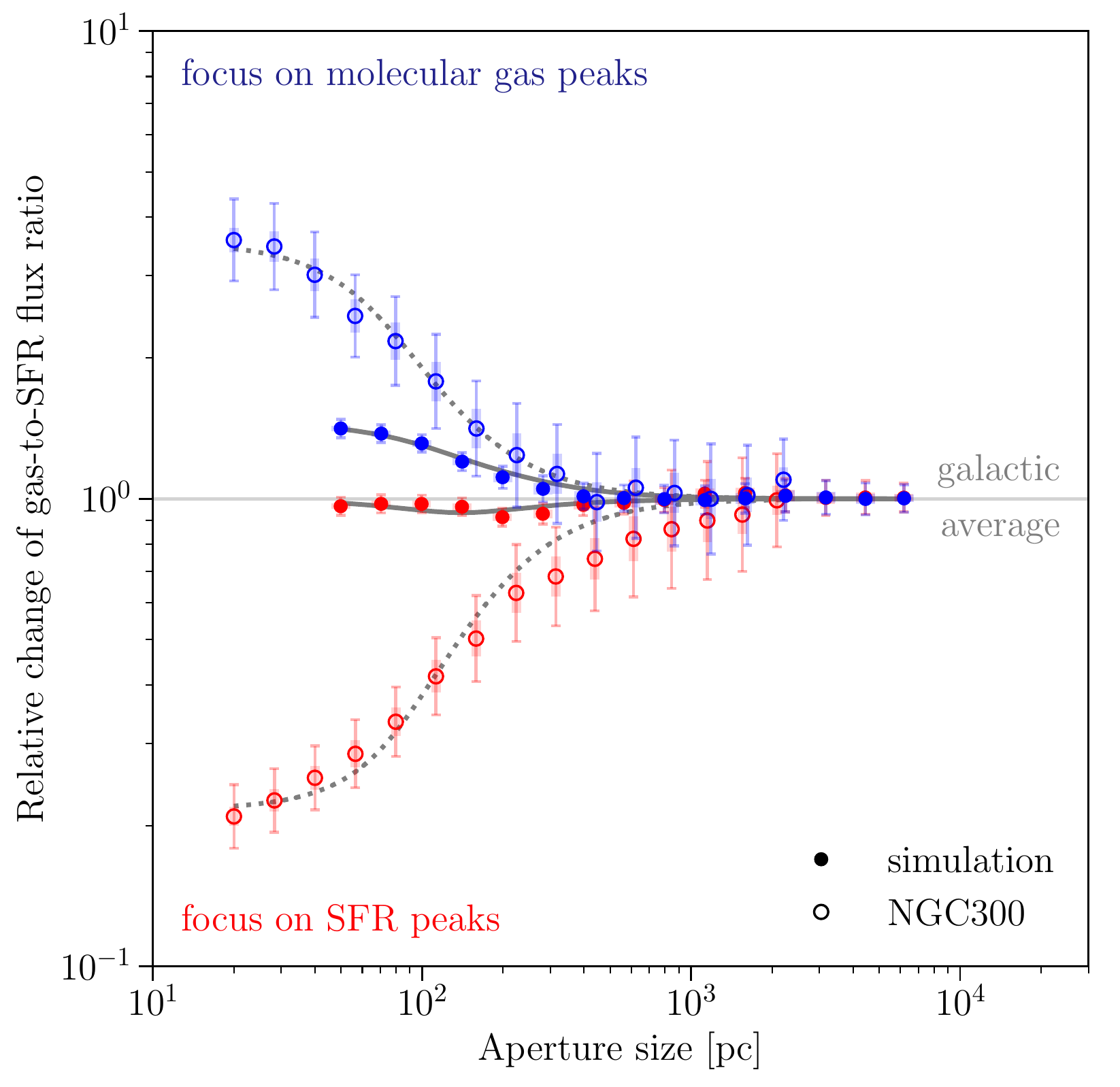}
\caption{Change of the gas-to-SFR flux ratio relative to the galactic average as a function of the aperture size, when focusing apertures on molecular gas peaks (top branch in blue; traced by synthetic CO(1-0) emission) or SFR peaks (bottom branch in red; traced by a mass surface density map of stars in the age bin 0-36~Myr). Data points with error bars show the measurements, whereas the lines show the best-fitting models. Thin error bars show the $1\sigma$ uncertainty for each individual data point, and the thick error bars show the effective uncertainty, accounting for the covariance between the data points. In the simulation (solid lines, closed symbols), the correlation between gas and star formation persists down to GMC ($<100$~pc) scales, whereas the real-Universe galaxy NGC300 (dotted lines, open symbols, combining CO and H$\alpha$ emission; taken from \citealt{kruijssen19}) exhibits a clear anti-correlation between gas and young stars on these scales. This shows that the prescription for stellar feedback used in the simulations is not sufficiently effective at disrupting the parent GMC.}
\label{fig:tuning_fork_diagram}
\end{figure}

In the previous subsections, we have shown that our Milky Way-like galaxy simulation reproduces all major observational constraints on the properties of the GMC population and the galaxy itself. We now subject the simulation to a new observational constraint on cloud-scale physics, the so-called `uncertainty principle for star formation' \citep{kruijssen14,KruijssenEtAl2018}, which characterizes the correlation or anti-correlation between tracers of molecular gas and star formation as a function of aperture size, by relating it to the evolutionary timeline of GMCs and star-forming regions. We show that, despite reproducing all other relevant quantities, the simulation does not reproduce the anti-correlation between gas and young stars that characterises observed galaxies on small spatial scales, and demonstrate that this is a direct result of ineffective stellar feedback on the GMC scale.

The `uncertainty principle' enables the empirical characterisation of the GMC lifecycle by placing apertures on peaks of gas or young stellar emission and measuring how the enclosed gas-to-young stellar flux ratios are elevated or suppressed, respectively, relative to the galactic average as the aperture size is changed. \autoref{fig:tuning_fork_diagram} shows this measurement for the simulation in comparison to observations of the nearby galaxy NGC300 \citep{kruijssen19}. On spatial scales $<500$~pc, the observations show an anti-correlation between gas and young stars (also see \citealt{SchrubaEtAl2010} and \citealt{chevance19}), manifesting itself as a split between the two branches focusing apertures on gas peaks and young stellar peaks towards smaller apertures. This characteristic `tuning fork' shape reflects the rapidity of evolutionary cycling between gas and young stars \citep{kruijssen19,KruijssenEtAl2018}. If GMCs remain co-spatial with young stars for many dynamical times (i.e.~much longer than the lifetimes of massive stars producing SFR tracers), then the gas-to-young stellar flux ratio on GMC scales should be similar to that observed on galactic scales, resulting in flat branches in \autoref{fig:tuning_fork_diagram}. Conversely, if GMCs undergo rapid lifecycles and become spatially separated from young stars after approximately one dynamical time (either because the GMCs are destroyed, or because they are displaced by stellar feedback), then GMCs and young stars should rarely coexist, causing the gas-to-young stellar flux ratio to strongly fluctuate on GMCs scales. Provided that the lifetime of one of the two phases (i.e.~either gas or young stars) is known \citep{haydon18} and any diffuse, inert emission has been removed from the emission maps \citep{hygate18}, a statistical model can be fitted to the diagram to infer the absolute evolutionary timeline \citep[see the dotted and solid lines in \autoref{fig:tuning_fork_diagram}]{KruijssenEtAl2018}.

\autoref{fig:tuning_fork_diagram} shows that the simulation is characterised by flat branches in this diagram, implying that the simulated GMCs form stars over long time-scales. Specifically, we measure an average GMC lifetime of $36_{-6}^{+4}$~Myr, which is in excellent agreement with the median GMC lifetime obtained from cloud tracking in \autoref{fig:cloud_lifetime}, as well as with the longest GMC lifetimes observed in nearby galaxies \citep[see][]{chevance19}. However, we measure a `feedback time-scale' of $23_{-1}^{+1}$~Myr. This time-scale indicates that after the appearance of the first massive stars, it takes more than 20~Myr before the parent GMC is disrupted. Given that the typical $\mathrm H\textsc{ii}$ region lifetime of a coeval stellar population is 4~Myr at solar metallicity \citep{haydon18}, this implies that star formation continues for many $\mathrm H\textsc{ii}$ region lifecycles in the simulated GMCs. This strongly contrasts with observed feedback time-scales in nearby galaxies, which are all $<10$~Myr \citep{chevance19}. 

\begin{figure}
\includegraphics[width=\hsize]{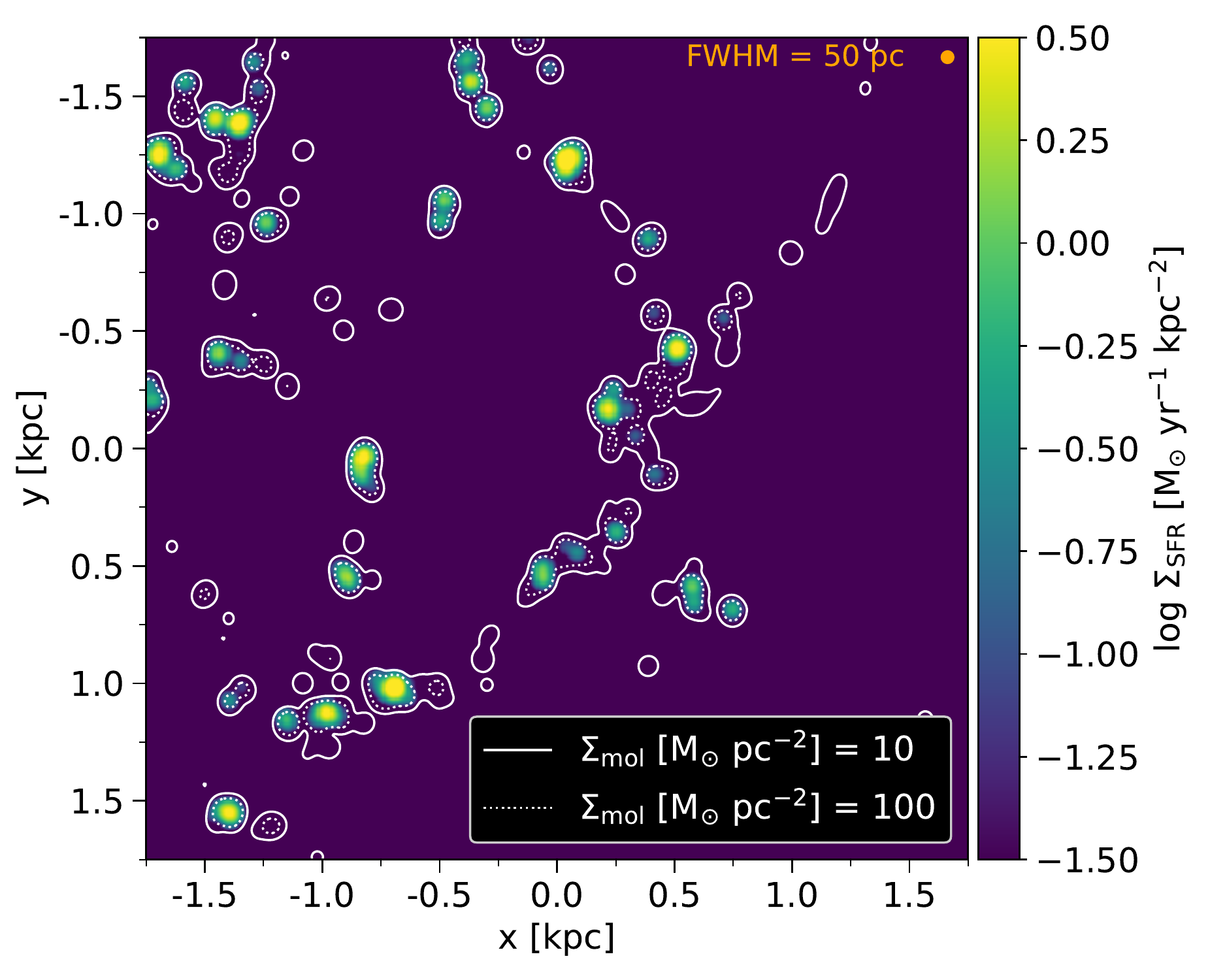}
\caption{Zoom-in image of gas tracer emission and young stellar tracer emission. Contours show CO $J = 1 \rightarrow 0$ line emission converted into the molecular gas surface density, while the colour map shows ionising luminosity converted to star formation rate per unit area. The region shown is a zoom in on a spot near $r = $ 6 kpc from the Galactic centre. Both the molecular gas and star formation maps have been convolved with a two-dimensional Gaussian kernel with a standard deviation of $\sigma =$ 21.2 pc, corresponding to a FWHM of 50 pc, shown as an orange circle at the top right corner.}
\label{fig:ionising_photons_with_CO_contour}
\end{figure}

We can see this effect visually in \autoref{fig:ionising_photons_with_CO_contour}, which
shows a zoom-in image of young stellar tracer emission overlaid with contours of molecular gas tracer emission. Almost all young stellar emission is associated with molecular gas, unlike in real galaxies where the two tracers de-correlate on $\lesssim 100$ pc scales. The obvious conclusion is that the stellar feedback in the simulations is incapable of disrupting GMCs as efficiently as in the real Universe. This shows that even when a simulation reproduces all of the standard properties of the GMC population and the galaxy at large, it may do so for the wrong reasons. The anti-correlation of gas and young stars on GMC scales provides a stronger test of the included feedback physics.

\section{Discussion and Conclusions}
\label{Conclusions}

In this paper we report the outcome of a detailed comparison between a state of the art hydrodynamic simulation of a Milky Way-like galaxy and a wide range of observations. The simulation includes all the bells and whistles commonly-found in modern high-resolution galaxy simulations: stochastic star formation, photoionisation heating, and individually-resolved SNe treated with a momentum-energy injection procedure. This simulation reproduces all macroscopic observational constraints for a Milky Way-like galaxy, including the total star formation rate, the partition of the ISM between warm and cold H\textsc{i} and molecular gas, radial profiles of velocity dispersion and Toomre $Q$ parameter, the kpc-scale Kennicutt-Schmidt relations measured separately for atomic, molecular, and total gas, and the mass spectrum, sizes, and virial parameters of GMCs. However, despite all these successes, we find that our simulation fails to reproduce the observed anti-correlation between molecular gas and star-formation tracers on $\lesssim 100$ pc scales, the so-called star formation uncertainty principle \citep{kruijssen14}. We find instead that almost all young stellar emission is associated with molecular tracer emission even on small scales, inconsistent with observations. This suggests that the observed scale-dependent anti-correlation between molecular gas and star formation represents a fundamental test for stellar feedback models in galaxy simulations, one that can fail for even state-of-the art simulations that reproduce the properties of galaxies on larger scales.\footnote{
We emphasise here that our focus is on simulations and models that seek to capture physics on $\lesssim 100$ pc scales. It is obviously not possible to deploy this test on, e.g., large-volume cosmological simulations that have resolutions of $\approx 1$ kpc, nor is it necessary to do so. Feedback recipes in simulations at this resolution are not intended to capture the internal structures of galaxies, just to produce roughly correct mass budgets for star formation and galactic winds.
}

One might suspect that this failure is a flaw in our mixed momentum-energy SN feedback recipe, which is only an approximation to the results that one would obtain by simulating at a resolution high enough to capture the Sedov-Taylor phase of SN remnant expansion. There is significant numerical evidence that real SN remnant expansion is more complex than simple recipes that adopt a fixed momentum budget per SN \citep[e.g.,][]{Kimm15a, GoldbaumKrumholzForbes2016, Hopkins18a}, particularly when SNe are clustered \citep{GentryEtAl2017, GentryEtAl2019, Hu2019}. However, this possibility is somewhat unlikely because observations suggest that star clusters are already gas-free before the first SNe occur, both in the Milky Way \citep{LongmoreEtAl2014} and in M83 \citep{HollyheadEtAl2015}. The spectra of clusters that are selected to be in the process of gas clearing frequently show Wolf-Rayet features, again suggesting that gas clearing occurs during an evolutionary phase that precedes the first SN \citep{SokalEtAl2016}. Finally, observed feedback destruction time-scales of entire GMCs are often shorter than the 3~Myr delay time associated with SN feedback \citep{kruijssen19,chevance19}. These observations disfavour SNe as the dominant mechanism for gas removal \textit{at the scale of star clusters and GMCs}, which is the physical cause of the small-scale de-correlation between gas and star formation.

We 
instead
trace the likely origin of our simulations' failure to reproduce the small-scale anti-correlation to our pre-SN feedback being insufficiently strong to disperse molecular clouds, or at least drive them away from star-forming regions. Our failure to reproduce the small-scale anti-correlation is likely not due to our having omitted a key process -- we include photoionisation heating, which both observational and theoretical arguments strongly suggest is the dominant feedback mechanism in galaxies like the Milky Way \citep{KrumholzEtAl2018}. Instead, we conjecture that our implementation of photoionisation feedback at thermal heating is inadequate given the resolution we achieve, and that the problem might be overcome by implementing it as momentum injection or a hybrid form of energy-momentum injection. There is an obvious analogy to SN feedback. While early galaxy simulations tended to implement SN feedback as simple thermal energy input, the discovery of the overcooling problem \citep{Katz1992} led to the development of alternate schemes, such as momentum or mixed-momentum energy approaches \citep[e.g.,][]{Kimm15a,GoldbaumKrumholzForbes2016,Hopkins18a}. The momentum budget associated with photoionised gas rocketing away from neutral surfaces is likely comparable to or larger than that of SNe \citep{KrumholzEtAl2018}, but at the resolution typical of even isolated galaxy simulations, this process cannot be resolved, leading to an underestimation of the effect. We intend to explore this possibility in future work.

\section*{Acknowledgements}

Simulations were carried out on the Cray XC30 at the Center for Computational Astrophysics (CfCA) of the National Astronomical Observatory of Japan and the National Computational Infrastructure (NCI), which is supported by the Australian Government. YF and MRK acknowledge support from the Australian Government through the Australian Research Council's \textit{Discovery Projects} and \textit{Future Fellowships} funding schemes (project DP160100695 and FT180100375). DTH is a fellow of the International Max Planck Research School for Astronomy and Cosmic Physics at the University of Heidelberg (IMPRS-HD). MC and JMDK gratefully acknowledge funding from the Deutsche Forschungsgemeinschaft (DFG) through an Emmy Noether Research Group (grant number KR4801/1-1). JMDK gratefully acknowledges funding from the European Research Council (ERC) under the European Union's Horizon 2020 research and innovation programme via the ERC Starting Grant MUSTANG (grant agreement number 714907). YF, DTH, MC, MRK, and JMDK acknowledge support from the Australia-Germany Joint Research Cooperation Scheme (UA-DAAD, grant number 57387355). Computations described in this work were performed using the publicly available \textsc{enzo} code (\citealt{BryanEtAl2014}; \url{http://enzo-project.org}), which is the product of a collaborative effort of many independent scientists from numerous institutions around the world. Their commitment to open science has helped make this work possible. We acknowledge extensive use of the \textsc{yt} package (\citealt{TurkEtAl2011}; \url{http://yt-project.org}) in analysing these results and the authors would like to thank the \textsc{yt} development team for their generous help.




\bibliographystyle{mnras}
\bibliography{reference} 




\appendix




\bsp	
\label{lastpage}
\end{document}